\documentclass[preprint]{aastex}
\usepackage{epsfig}

\newcommand {\be} {\begin{equation}}
\newcommand {\ee} {\end{equation}}

\def\te{T_{\rm e}}

\begin{document}

\title{Timing Properties of Magnetars}
 
 \author{Feryal \"Ozel\altaffilmark{1}} \affil{Harvard-Smithsonian Center for
 Astrophysics and Physics Department, Harvard University, \\ 60 Garden
 St., Cambridge, MA 02138; fozel@cfa.harvard.edu }
\altaffiltext{1}{Present address: Institute for Advanced Study, Einstein Dr.,
       Princeton, NJ 08540}

\begin{abstract}

We study the pulse morphologies and pulse amplitudes of thermally
emitting neutron stars with ultrastrong magnetic fields.  The beaming
of the radiation emerging from a magnetar was recently shown to be
predominantly non-radial, with a small pencil and a broad fan
component. We show that the combination of this radiation pattern with
the effects of strong lensing in the gravitational field of the
neutron star yields pulse profiles that show a qualitatively different
behavior compared to that of the radially-peaked beaming patterns
explored previously.  Specifically, we find that: {\em (i)} the pulse
profiles of magnetars with a single hot emission region on their
surface exhibit $1-2$ peaks, whereas those with an antipodal emission
geometry have $1-4$ peaks, depending on the neutron star compactness,
the observer's viewing angle, and the size of the hot regions; {\em
(ii)} the energy dependence of the beaming pattern may give rise to
weakly or strongly energy-dependent pulse profiles and may introduce
phase lags between different energy bands; {\em (iii)} the non-radial
beaming pattern can give rise to high pulsed fractions even for very
relativistic neutron stars; {\em (iv)} the pulsed fraction may not
vary monotonically with neutron star compactness; {\em (v)} the pulsed
fraction does not decrease monotonically with the size of the emitting
region; {\em (vi)} the pulsed fraction from a neutron star with a
single hot pole has, in general, a very weak energy dependence, in
contrast to the case of an antipodal geometry.  Comparison of these
results to the observed properties of anomalous X-ray pulsars strongly
suggests that they are neutron stars with a single hot region of
ultrastrong magnetic field.
\end{abstract}
 
\keywords{radiation mechanisms:thermal --- stars:magnetic fields --- 
 stars:neutron --- X-rays:stars}

\section{Introduction}

Anomalous X-ray Pulsars (AXPs) and Soft Gamma-ray Repeaters (SGRs) are
two classes of intriguing objects that have challenged the standard
paradigm of young neutron stars. Particularly in the case of AXPs, two
types of models have been proposed to explain their spectral and
timing properties: conventional accretion models modified to account
for the absence of an observable donor star (van Paradijs, Taam, \&
van den Heuvel 1995; Chatterjee, Hernquist, \& Narayan 2000; Alpar
2001) and magnetar models with several different mechanisms for
powering the X-ray emission (Thompson \& Duncan 1996; Heyl \&
Hernquist 1998). In all these magnetar models, a significant fraction
of the X-ray emission is thought to originate at the neutron star
surface.

Recently, there has been significant progress in the calculation of
the surface emission properties of ultramagnetized neutron stars and
their application to AXPs (\"Ozel 2001; \"Ozel, Psaltis, \& Kaspi
2001; see also Ho \& Lai 2001; Zane et al.\ 2001). \"Ozel (2001)
showed that the angle dependence of surface radiation in such strong
fields has both a pencil and a fan component, with the latter
dominating in the X-ray range. Such a beaming pattern can give rise to
a variety of pulse profiles and timing properties that have not been
explored to date. Furthermore, the general relativistic bending of
photon trajectories in the strong gravitational field of the neutron
star significantly affects the observed properties of emission
originating at the stellar surface. This is in contrast to
rotation-powered pulsars whose radio emission originates at the light
cylinder far above the stellar surface, where the self lensing by the
neutron star is weak. Therefore, in order to compare in detail the
models with the current and future observations of AXPs, it is now
necessary to carry out a complete study of the observable properties
of magnetars taking all the above effects into account.

The advent of X-ray telescopes with good timing and energy resolution
as well as broad spectral coverage has produced high-quality data on
AXPs in the recent years, making such a study timely. Indeed,
observations with {\em ASCA} and {\em BeppoSAX} have yielded good
broad-band spectra of AXPs as well as a measure of the flux variations
during a pulse cycle in multiple energy bands (e.g., White et al.\
1996; Oosterbroek et al.\ 1998). Moreover, the superb timing
resolution of the {\em Rossi X-ray Timing Explorer (RXTE)} made
possible the phase connection of pulse cycles over several years and
hence produced detailed energy-dependent pulse profiles (Gavriil \&
Kaspi 2002). Finally, the grating spectrographs onboard the {\em
Chandra} and {\em XMM-Newton} observatories extend the energy range of
current observations towards the soft X-rays in addition to allowing
for phase-resolved spectroscopy and searches for spectral lines (e.g.,
Patel et al.\ 2001; Juett et al.\ 2002; Tiengo et al.\ 2002). It has
been argued that the combination of the pulse profiles, the amplitude
of pulsations, and the pulse-phase resolved spectral features observed
with these instruments provide the most stringent constraints on the
models of AXPs (\"Ozel et al. 2001).

In this paper we explore the pulse profiles expected from a strongly
magnetized neutron star emitting thermally from its hot surface.  In
detail, we focus on the dependence of the pulse profiles on the
emission geometry, the orientation of the observer, the photon energy,
and the compactness of the neutron star. We also discuss the
implications of our results for magnetar models of AXPs.

\section{The Model}

We consider the emission from a hot ultramagnetized neutron star. In
order to determine the properties of its surface emission, we carry
out radiative transfer calculations in strong fields and construct
model atmospheres in radiative equilibrium. We assume that the stellar
atmosphere is a completely ionized electron-proton plasma in
plane-parallel geometry, owing to its negligible thickness compared to
the stellar radius. We take the magnetic field to be orthogonal to the
surface, motivated by the fact that the emission most probably
originates at the magnetic poles. We consider fully angle- and
energy-dependent emission, absorption, and scattering processes in the
two photon polarization modes of a magnetized plasma and include the
effects of vacuum as well as of the plasma on photon propagation.  We
neglect the effects of the neutron star magnetosphere which may affect
the photon interactions above the stellar atmosphere. A detailed
description of the model as well as the surface emission properties
are given in \"Ozel (2001).

There are two modifications in the present calculation regarding the
treatment of vacuum polarization, which have a weak effect on the
results. First, for magnetic field strengths above the critical value
$B_{\rm cr} = 4.414 \times 10^{13}$~G, we modify the interaction cross
sections following Tsai \& Erber (1975). Second, we employ an adaptive
mesh technique to resolve the sharp features associated with the
vacuum polarization resonance instead of the saturation technique
presented in \"Ozel (2001). The details of these improvements as well
as the effects of the proton cyclotron resonance on the emerging
spectra are discussed in \"Ozel (2002).

We consider two emission geometries: {\em (i)} a single hot pole that
corresponds to, e.g., an isolated region of enhanced magnetic
activity, and {\em (ii)} two antipodal hot regions corresponding to,
e.g., two magnetic poles. In both cases, the temperature is taken to
be uniform across the hot region and is specified by an effective
temperature $\te$, while the rest of the neutron star is assumed not
to radiate.  We allow for variations of the angular size of the
emitting region $\rho$ as well as of the orientation angles $\alpha$,
which specifies the location of the magnetic poles with respect to the
rotation axis on the neutron star surface, and $\beta$, which
determines the position of the observer with respect to the same axis.
For most of the calculations presented here, we fix the magnetic field
strength to $B=10^{15}$~G. We also use an effective temperature
$\te=0.5$~keV, which yields spectra with color temperatures at
infinity of $\approx 0.5-0.6$~keV at this field strength, appropriate
for the spectral characteristics of AXPs (see \"Ozel et al.\ 2001).

In calculating the observables at infinity, we take into account the
gravitational lensing of the surface emission by the neutron star. We
use the photon trajectories given by the Schwarzschild metric, which
provides a good approximation for slowly rotating neutron stars such
as AXPs, and follow the method described by Pechenick, Ftaclas, \&
Cohen (1983) to compute light curves.  We vary the neutron star
relativity parameter $p=R c^2/2 G M$, where $R$ and $M$ denote the
radius and the mass of the neutron star, respectively.  We consider
the range $2 \le p \le 4$, which is allowed by current equations of
state for a wide range of neutron star masses (see, e.g., Cardall,
Prakash, \& Lattimer 2001). Note that the magnetic pressure becomes
important in hydrostatic balance for very large ($B \sim 10^{17}$~G)
field strengths and may lead to larger neutron star radii in this
regime. However, because the magnetic field does not significantly
affect the equation of state at the lower values considered here, we
take the often employed value of $p=2.5$ corresponding to a $1.4
M_\odot$ and $10$~km neutron star as the fiducial value when varying
the other model parameters.

\subsection{The Beaming of Surface Emission}

The pulse profiles are determined primarily by the angle dependence
(beaming) of the radiation emerging from the neutron star surface and
the general relativistic bending of photon trajectories in strong
gravitational fields. In the ultrastrong magnetic fields that we
consider here, the beaming is qualitatively different than that of
weakly-magnetic, non-magnetic, or accreting neutron stars and is
predominantly non-radial. A detailed discussion of the beaming of
surface radiation from a magnetar is given in \"Ozel (2001). Here, we
only summarize the results to facilitate the discussion of the pulse
profiles.

\begin{figure}[t]
\centerline{ \psfig{file=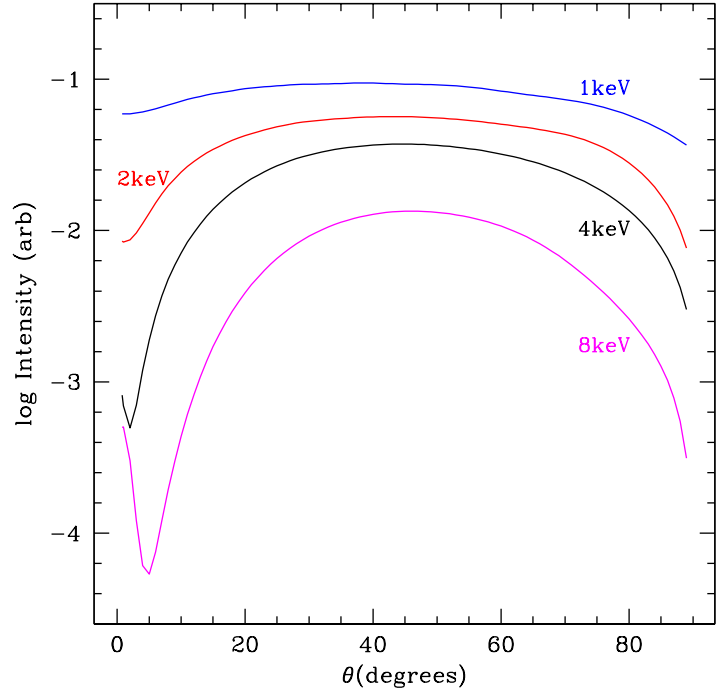,angle=0,width=9.5truecm} }
\figcaption[]{The angle dependence (beaming) of radiation emerging
from the surface of a neutron star with $B=10^{15}$~G, at photon
energies of $1, 2, 4$ and 8~keV. \label{Fig:beaming}}
\end{figure}

Figure~\ref{Fig:beaming} shows the beaming of radiation emerging from
the surface of a neutron star with $B=10^{15}$~G and $\te=0.5$~keV, at
photon energies $E=1, 2, 4$, and 8~keV. The abscissa $\theta$ denotes
the angle between the direction of the magnetic field and the
direction of propagation. In all cases, the radiation pattern can be
described by two components: a radial beam confined to $\theta
\lesssim $~few~degrees and a dominant wide ``fan'' beam which has a
peak at $\theta_{\rm p}\approx 40^\circ-60^\circ$. The radial beam
carries little flux and is observationally insignificant because of
its confinement to very small angles. Therefore, the pulse profiles,
and in particular the number and location of the peaks, carry
signatures of the fan beams. Note that when viewed along the magnetic
axis on the neutron star surface, this beaming pattern represents a
hollow cone of radiation whose opening angle and thickness vary with
photon energy.

The beaming of radiation shown in Figure~\ref{Fig:beaming} varies
weakly with magnetic field in the range $10^{14}-10^{15}$~G.
Therefore, the pulse profiles do not change significantly in this
range of field strengths.

\begin{figure}[t]
 \centerline{ \psfig{file=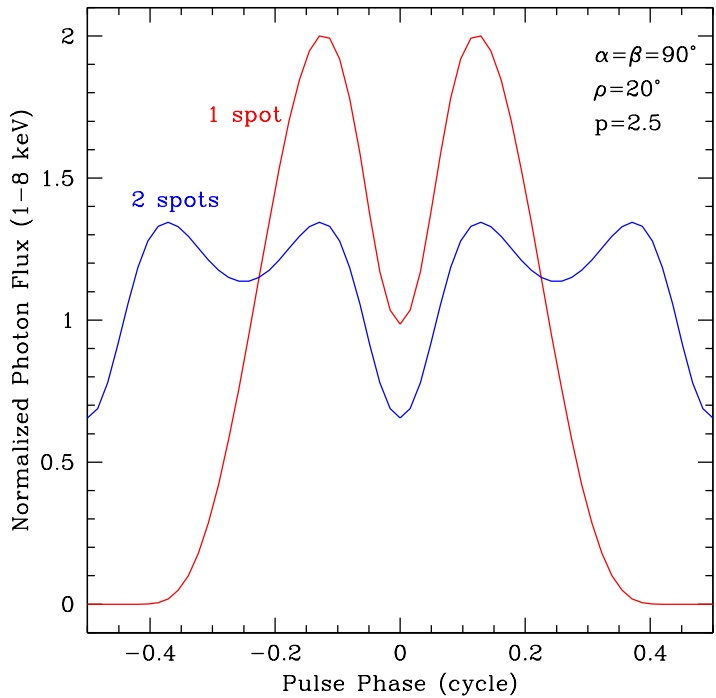,angle=0,width=9.5truecm} }
\figcaption[]{Sample $1-8$~keV pulse profiles for a neutron star with
a single or two antipodal hot regions. The former case shows two and
the latter four peaks per pulse cycle when the observer traverses the
emission cones at the center. \label{Fig:spots}}
\end{figure}

\section{Bolometric Pulse Profiles}

In this section, we present energy-integrated pulse profiles of
magnetars emitting thermally from their surface. We choose a $1-8$~keV
photon energy range, which is accessible with high sensitivity with
the current detectors and is thus relevant for comparison with the
present data on AXPs. For simplicity, we refer to quantities computed
in this energy range as bolometric.  We focus on the number and
location of the peaks in the pulse profile, as well as on the
peak-to-peak changes of the flux through a pulse cycle. We will
consider the photon-energy dependence of the profiles and further
quantify the modulation (through the use of the pulsed fraction) in
\S4 and \S5, respectively.

\begin{figure}[t]
\centerline{ \psfig{file=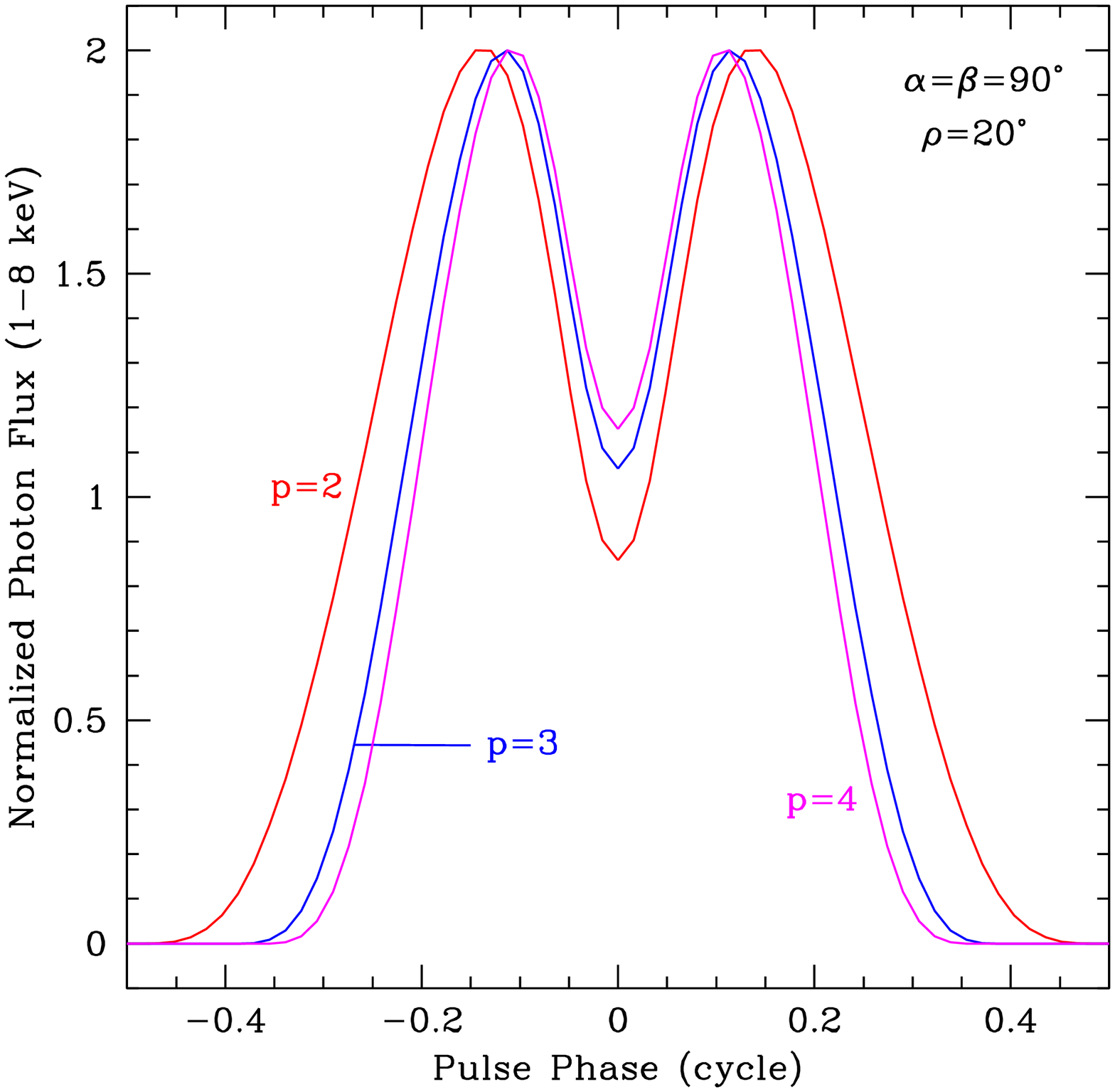,angle=0,width=8.3truecm} }
\centerline{ \psfig{file=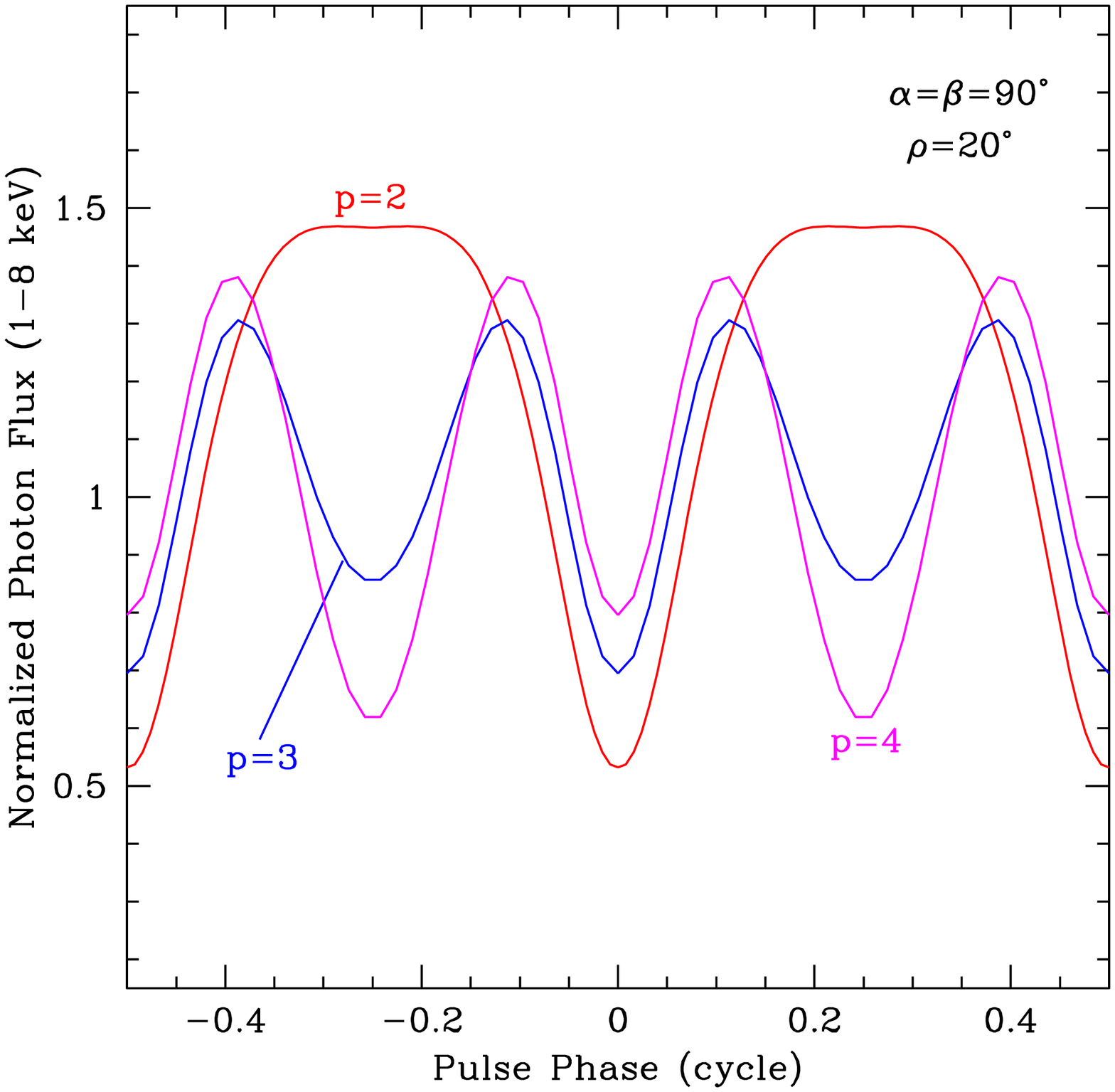,angle=0,width=8.3truecm} }
\figcaption[]{The $1-8$~keV pulse profiles for varying neutron star
compactness $R/2M=2, 3$ and~4. The neutron star has a magnetic field
of $10^{15}$~G and an effective temperature $T_{\rm eff}=0.5$~keV.
The top panel shows the result for a single hot pole while the bottom
panel corresponds to an antipodal emission geometry, with an angular
size of the hot regions $\rho=15^\circ$ in both cases. The two
orientation angles $\alpha$ and $\beta$ are taken to be $90^\circ$
(orthogonal rotator). \label{Fig:ppbol_p}}
\end{figure}

\begin{figure}[t]
\centerline{ \psfig{file=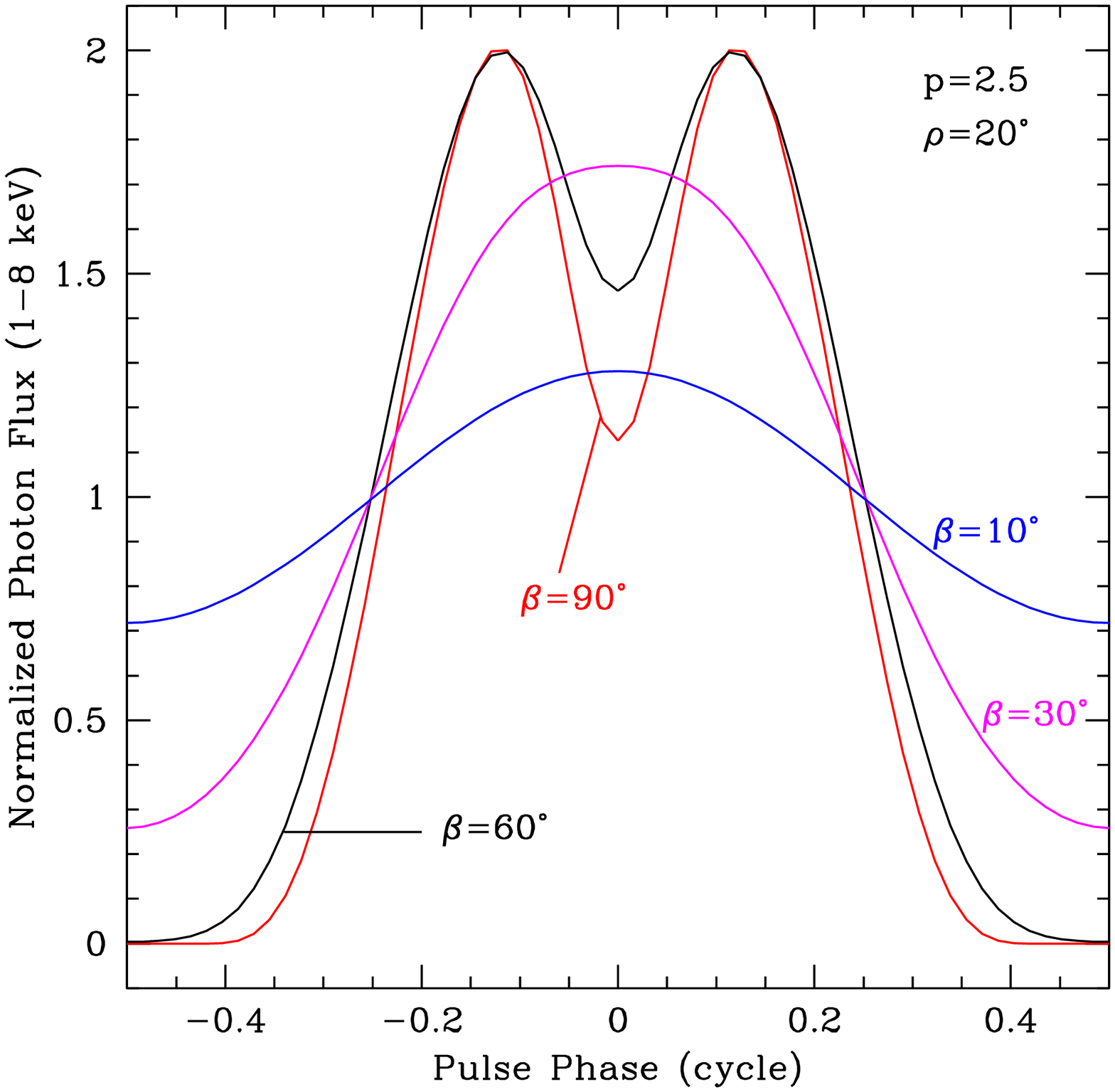,angle=0,width=8.3truecm} }
\centerline{ \psfig{file=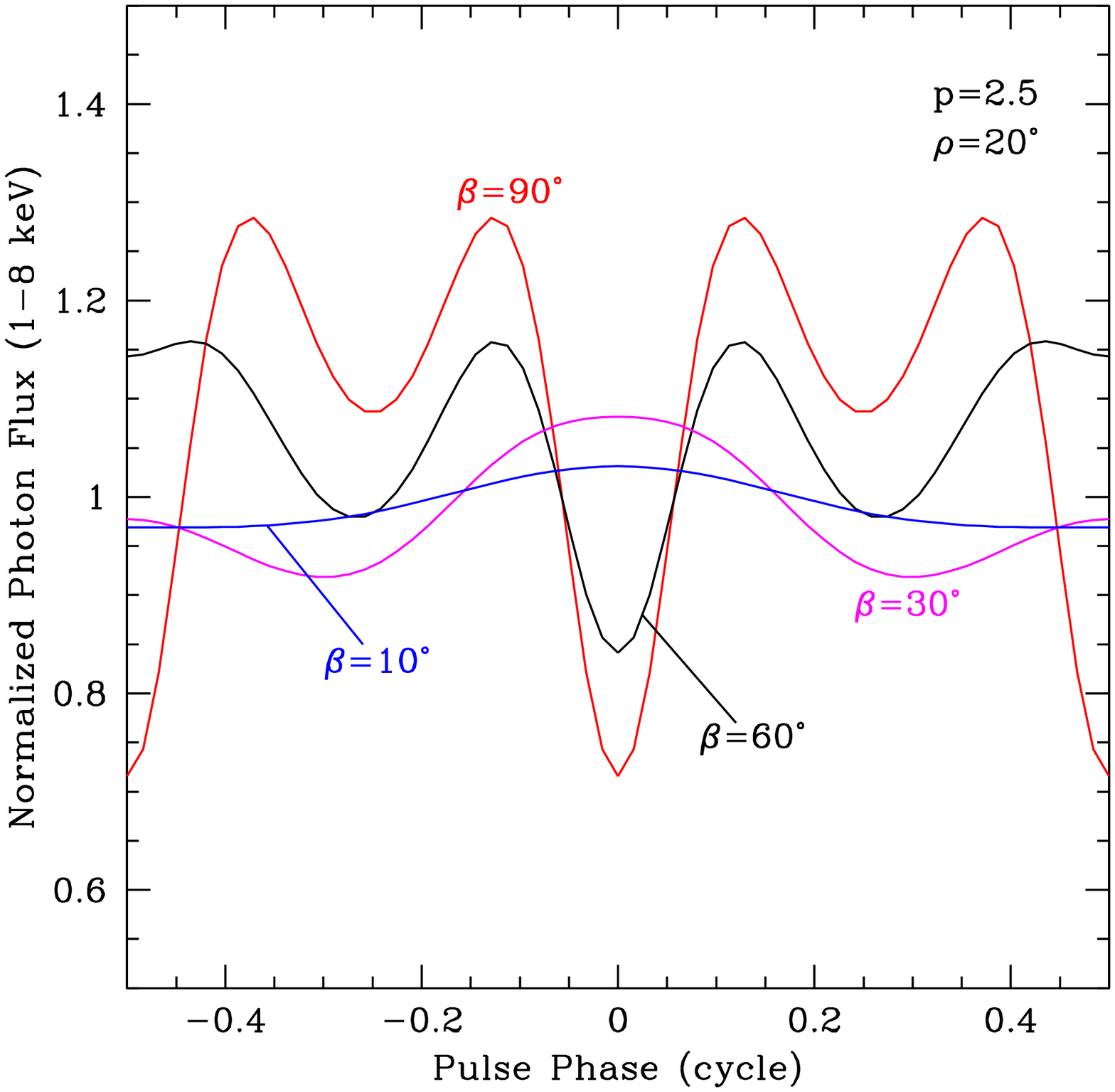,angle=0,width=8.3truecm} }
\figcaption[]{The $1-8$~keV pulse profiles for different values of the
observer angle $\beta$. Here, $\alpha=80^\circ$ while the field
strength and effective temperature are as in Figure~3. The top panel
shows $1-2$ peaks arising from a single pole and the bottom panel
displays the $1-4$ peaks that can be obtained in the antipodal
emission geometry. \label{Fig:ppbol_b}}
\end{figure}

Figures~\ref{Fig:spots}--\ref{Fig:ppbol_b} show bolometric pulse
profiles from a $B=10^{15}$~G neutron star with $\te=0.5$~keV. The
ordinate shows the photon flux normalized to the average flux through
a pulse cycle, while in the abscissa $\phi=0$ corresponds to the phase
when the observer is aligned with the center of one pole. The pulse
profiles shown in these figures are determined primarily by the
beaming of radiation on the neutron star surface as well as by the
strong gravitational lensing by the neutron star. In three dimensions,
the non-radial beaming pattern (Fig.~\ref{Fig:beaming}) corresponds to
a bright hollow cone of radiation with angular size $\sim \theta_p$
and thickness $\approx 30^\circ$ originating from every surface
element on the neutron star. Thus, in the absence of gravitational
lensing and for an observer traversing the diameter of a small
emitting region, two peaks appear per hot pole, at a phase that
corresponds to the location of the peaks in the beaming pattern. Thus,
the phase difference between the peaks originating from one pole is
approximately $2 \theta_p$.  Figure~\ref{Fig:spots} shows the pulse
profiles for this simplest case for the two emission geometries
discussed in \S2: a single pole gives rise to 2 peaks while an
antipodal emission geometry produces 4 peaks.

As the above discussion suggests, in more realistic cases, the
properties of the pulse profiles are determined not only by the number
of hot regions on the stellar surface but also by {\em (i)} general
relativistic effects, {\em (ii)} the orientation angles that describe
the positions of the emitting regions and of the observer with respect
to the rotation axis and {\em (iii)} the sizes of the emitting
regions. We now explore each of these effects in detail, investigating
the resulting number and separation of the peaks in the pulse profiles
and the peak-to-peak changes of the flux during a pulse phase.

\subsection{The Effect of the Gravitational Lensing by the Neutron Star}

General relativistic effects spread out and smear the beaming
patterns, effectively changing the observed opening angles and
thicknesses of the hollow radiation cones. In general, self-lensing by
the neutron star makes more of the stellar surface visible to the
observer at any pulse phase and leads to a suppression of the
modulation of the observed lightcurves. In some cases, however,
non-radial beams can be focused away from the surface normal by the
gravitational lensing in a way that the flux at particular pulse
phases can be enhanced. This happens when cones of radiation
originating from different parts of one hot region or from two
antipodal hot regions are deflected towards each other and smeared so
strongly by the stellar gravity that they overlap. This effect can
also change the number of peaks in the pulse profile. Thus, the
general relativistic effects in conjunction with a predominantly
non-radial radiation pattern (fan beaming) give rise to qualitatively
different results, affecting the number and separation of the peaks in
the pulse profile as well as the overall flux modulation.

Figure~\ref{Fig:ppbol_p} shows the effect of varying the relativity
parameter $p$ on the pulse profiles.  The $p=4$ case represents a
large neutron star where the curvature of photon paths is small, and
the result is nearly Newtonian with 4~peaks located roughly at phases
$\phi= \pm \theta_{\rm p}$ and $\pi \pm \theta_{\rm p}$. As $p$
decreases, i.e., as the neutron star becomes more relativistic,
gravitational lensing becomes stronger. The beams detected by an
observer at infinity are correspondingly defocused and appear with
larger angular sizes. Therefore, the observed peaks in the pulse
profile get broader and move to phases $\phi > \theta_p$. Another
consequence is that more flux reaches the observer during the
interpulses, as in the case of $p=3$ shown in
Figure~\ref{Fig:ppbol_p}.

As the neutron star becomes even more relativistic, pulse profiles in
the antipodal emission case go through a significant morphological
change because the cones of emission from the antipodal hot regions
start to overlap and produce an enhanced signal at $\phi = \pm
\pi/2$. When the beams combine in this way, the number of peaks in the
pulse profile is reduced and the overall modulation is increased
(Fig.~\ref{Fig:ppbol_p}). Note that for small values of $p$, the same
effect can be seen in the emission from a single pole, when its area
is large enough for the beams from two sides of the hot region to
combine (also see \S 3.3 and Fig.~\ref{Fig:ppsize} below). In that
case, the number of peaks in the pulse phase is reduced from two to
one. To summarize, the harmonic structure and modulation of the pulse
profiles are directly determined by the neutron star compactness, with
a sharp transition in the properties of the profiles at small
relativity parameters $p$.  For most values of $\alpha$, $\beta$ and
$\rho$, this transition in the pulse profiles occurs at $p \simeq
2.5$.

\subsection{The Effect of Geometry: Orientation Angles}

The orientation angles $\alpha$ and $\beta$ determine two aspects of
the pulse profiles. First, because they indicate the position of the
poles and of the observer with respect to the rotation axis, they
control the overall modulation of the lightcurves, ranging from
no modulation at $\alpha=0$ or $\beta=0$ to maximum modulation at
$\alpha=\beta=90^\circ$. Second, their relative values determine the
chord that the observer's path cuts through the cone of emission, and
thus the observed number and separation of peaks in the pulse profile.

Figure~\ref{Fig:ppbol_b} shows the pulse profiles for different values
of the observer's angle $\beta$. Here, we fix the position angle
$\alpha$ to $80^\circ$ and vary $\beta$ between $10^\circ$ and
$90^\circ$ for an emitting region of size $\rho=20^\circ$ and a
relativity parameter $p=2.5$. The striking result is that the number
of peaks can vary between $1-2$ for the single pole and between $1-4$
for the antipodal emission geometry, determined by the number of times
the observer intersects the cones of emission. However, unlike the
case of neutron star compactness, which also affects the observed
number of peaks, the reduction from four peaks to one due to different
observer's angles is usually accompanied by a suppression in the
pulsed fraction.

\begin{figure}[t]
\centerline{ \psfig{file=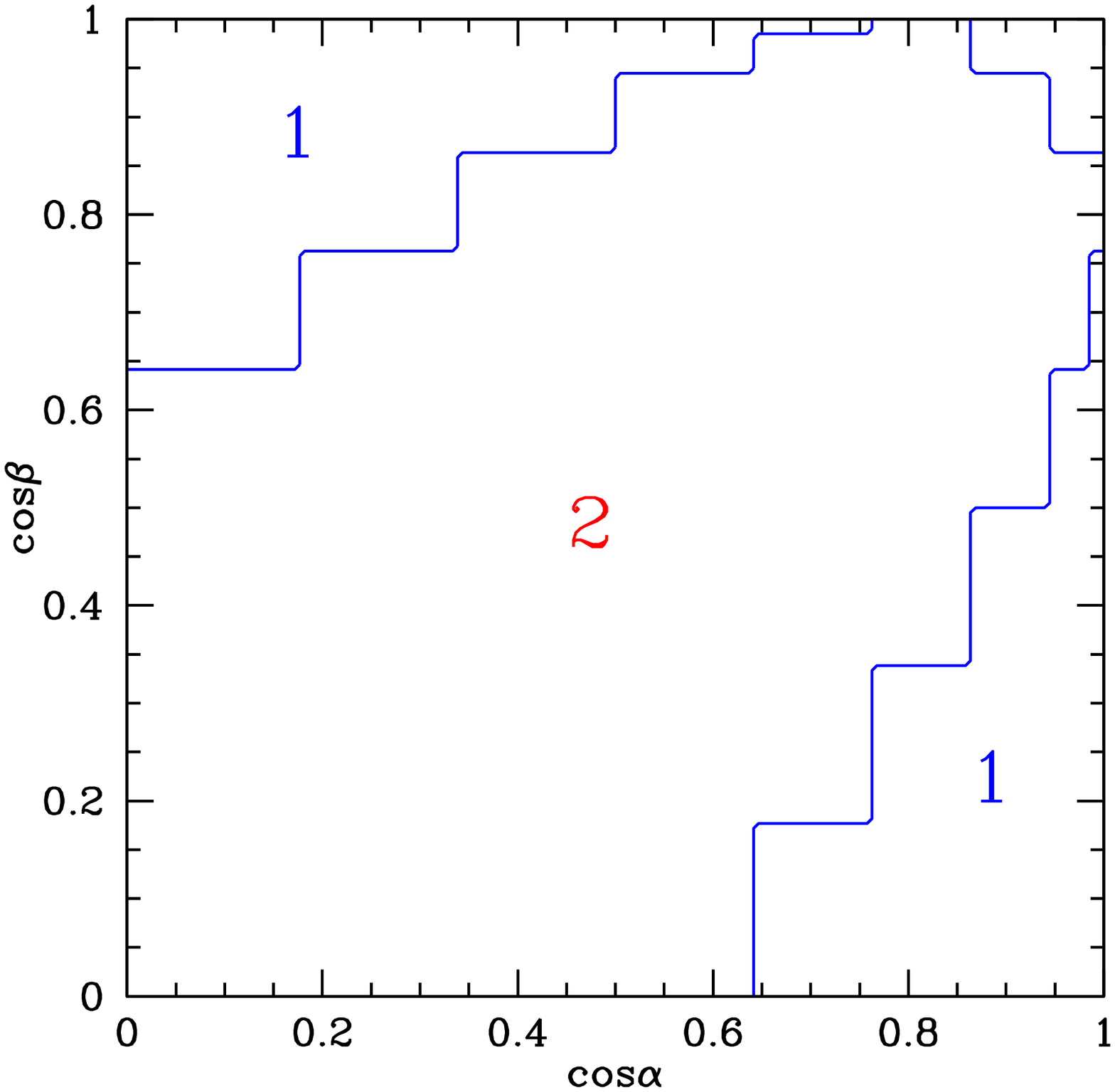,angle=0,width=8.2truecm} }
\centerline{ \psfig{file=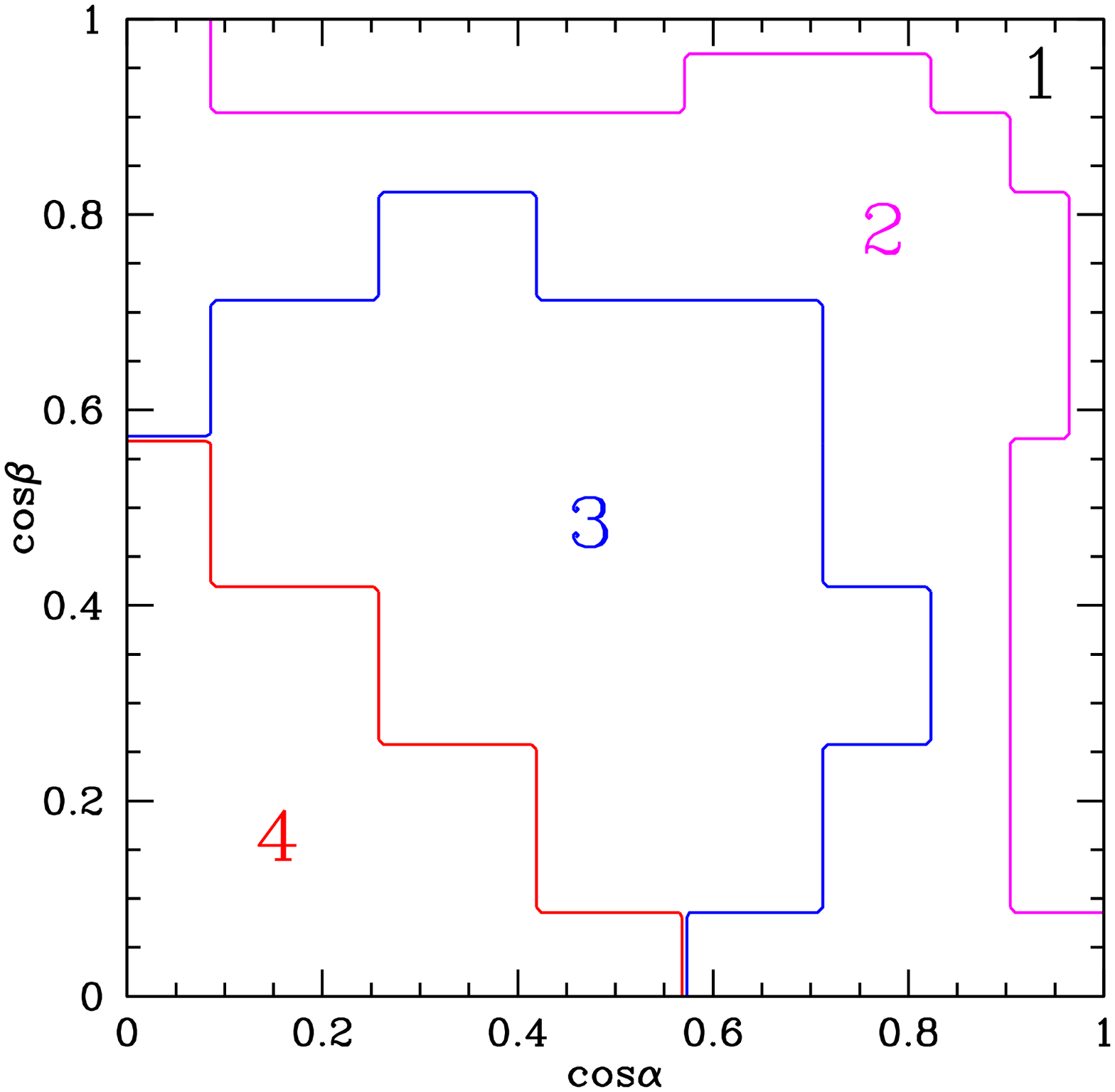,angle=0,width=8.2truecm} }
\figcaption[]{The contours show the number of peaks in the pulse
profile as a function of the cosine of the orientation angles $\alpha$
and $\beta$ for a single pole (top panel) and the antipodal geometry
(bottom panel).  All other model parameters are as in
Figure~4. \label{Fig:peaks_ab}}
\end{figure}

In Figure~\ref{Fig:peaks_ab}, we identify the regions in the ($\cos
\alpha, \cos \beta$) parameter space that produce different number of
peaks in the observed pulse profiles, the top panel corresponding to a
single pole and the bottom panel to the antipodal emission geometry.
For a random distribution of orientation angles, the areas covered by
each region is directly proportional to the probability of observing a
system with a given number of peaks in its pulse profile.

\subsection{The Effect of the Size of the Hot Regions}

\begin{figure}[h]
\centerline{ \psfig{file=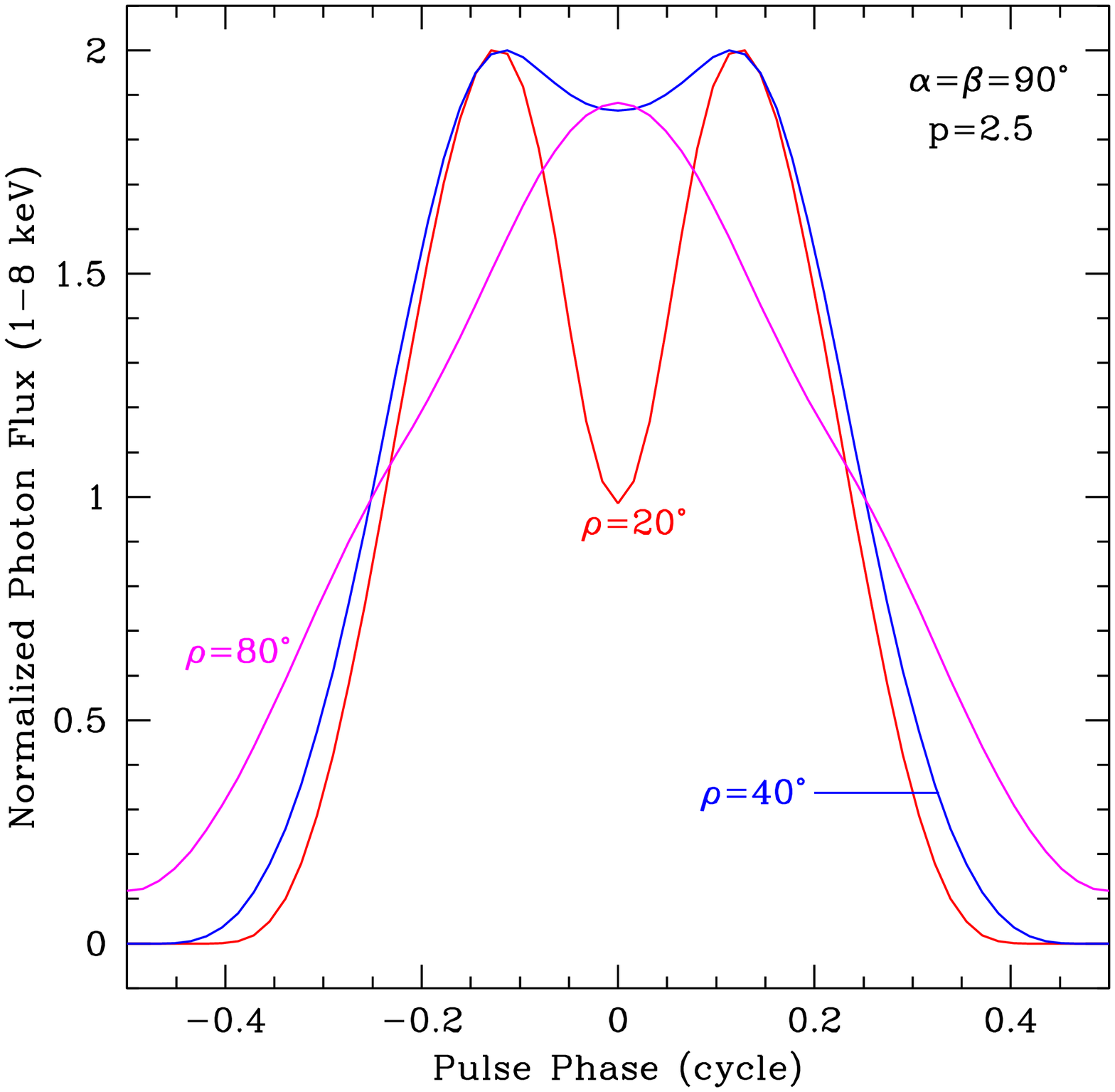,angle=0,width=8.2truecm} }
\centerline{ \psfig{file=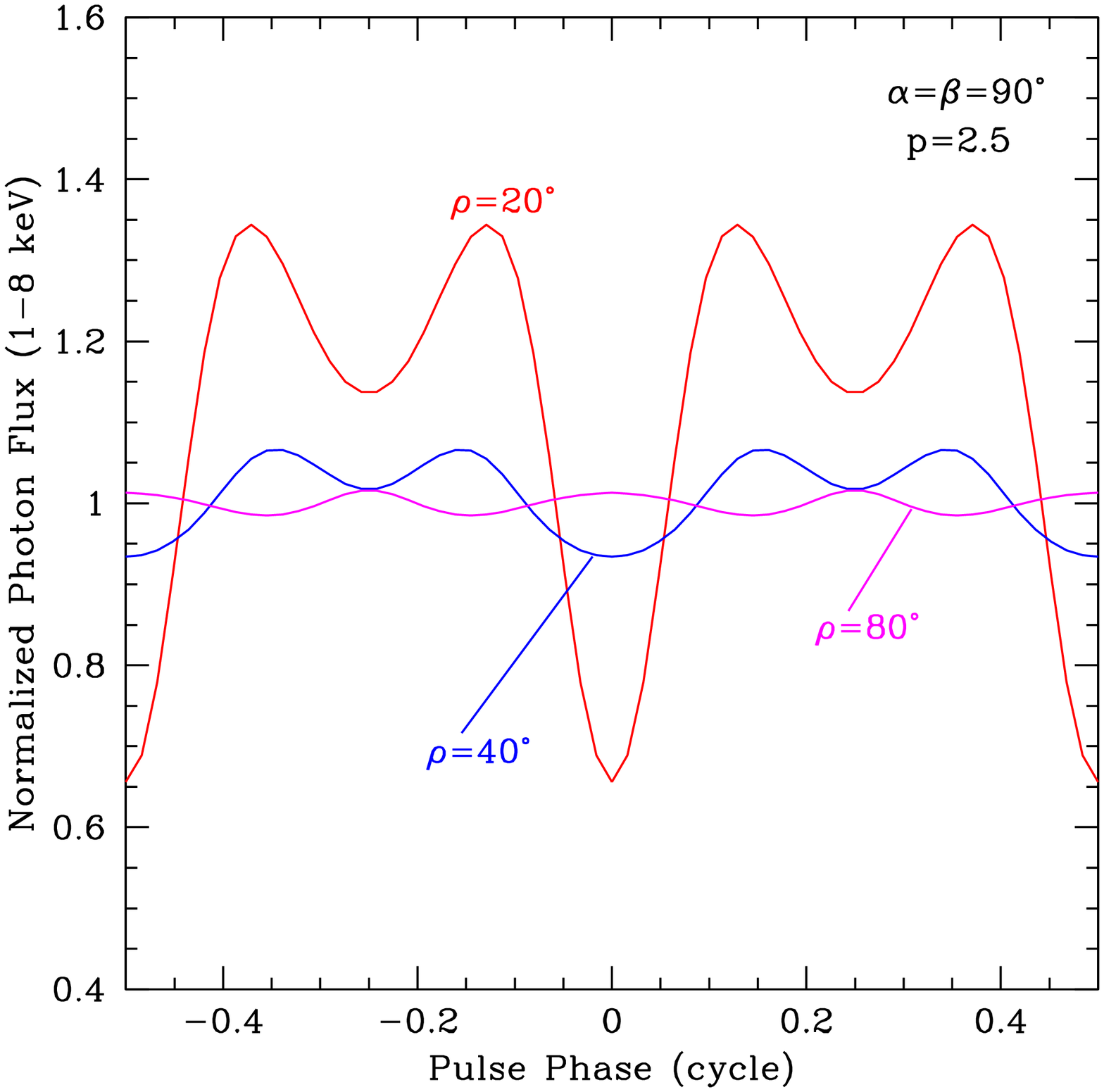,angle=0,width=8.2truecm} }
\figcaption[]{The dependence of the pulse profiles on the size of the
emitting region for (top panel) a single pole and (bottom panel) two
antipodal hot regions on the neutron star. All other model parameters
are the same as before. \label{Fig:ppsize}}
\end{figure}

The last parameter that affects significantly both the number of peaks
and the modulation of the pulse profiles is the area covered by the
hot regions on the stellar surface.  Figure~\ref{Fig:ppsize} shows the
dependence of the pulse profiles on the size of the emitting region
for both the single and antipodal emission geometries. As discussed
earlier, for large emitting regions, the observer receives the most
flux at a phase when the line of sight coincides with the center of
the pole because the non-radial emission from the outer edges of the
pole is directed towards the center. This enhancement reduces the
number of peaks to one in the case of the single pole while it shifts
the phase of the peaks by $\approx \pi/4$ in the antipodal emission
case.

\section{Photon-Energy Dependence of the Pulse Profiles and Phase Lags}

In this section, we investigate the dependence of pulse profiles on
photon energy in an energy range of interest for observations of AXPs.
The pulse profiles of strongly magnetized neutron stars are expected
in general to have a dependence on photon energy because of the
different beaming of the emerging radiation at different photon
energies. However, the properties of the pulse profiles in different
energy bands are also affected strongly by the number and sizes of the
emitting regions as well as by the orientation angles $\alpha$ and
$\beta$, as we show below. Following Gavriil \& Kaspi (2002), we
choose the energy ranges $1-2$, $2-4$, and $6-8$~keV for this study.

Because the photon-energy dependence of the pulse profiles changes
with all of the parameters that specify the geometry of the surface
temperature distribution and the position of the observer, it is not
feasible to illustrate the pulse profiles for the entire space spanned
by these parameters. Instead, we show in Figures~$7-10$ some
representative cases. In particular, the examples include cases where
the dependence of the pulse profiles and/or pulse amplitudes on photon
energy is either very weak or rather strong. We discuss the range of
parameters for which each behavior can be observed.

\vspace{-0.2cm}
\begin{figure}[t]
\centerline{ \psfig{file=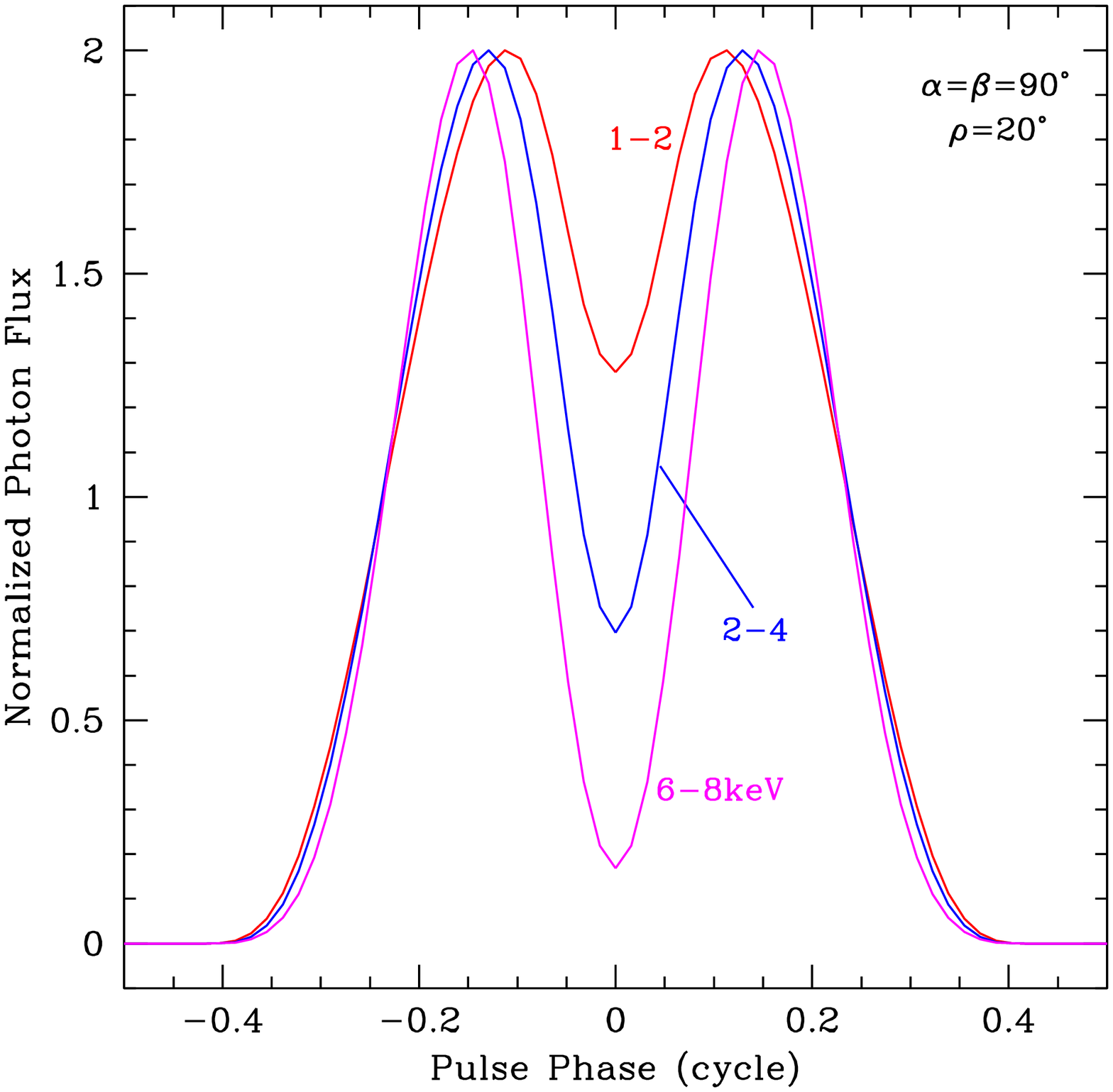,angle=0,width=8.2truecm} }
\centerline{ \psfig{file=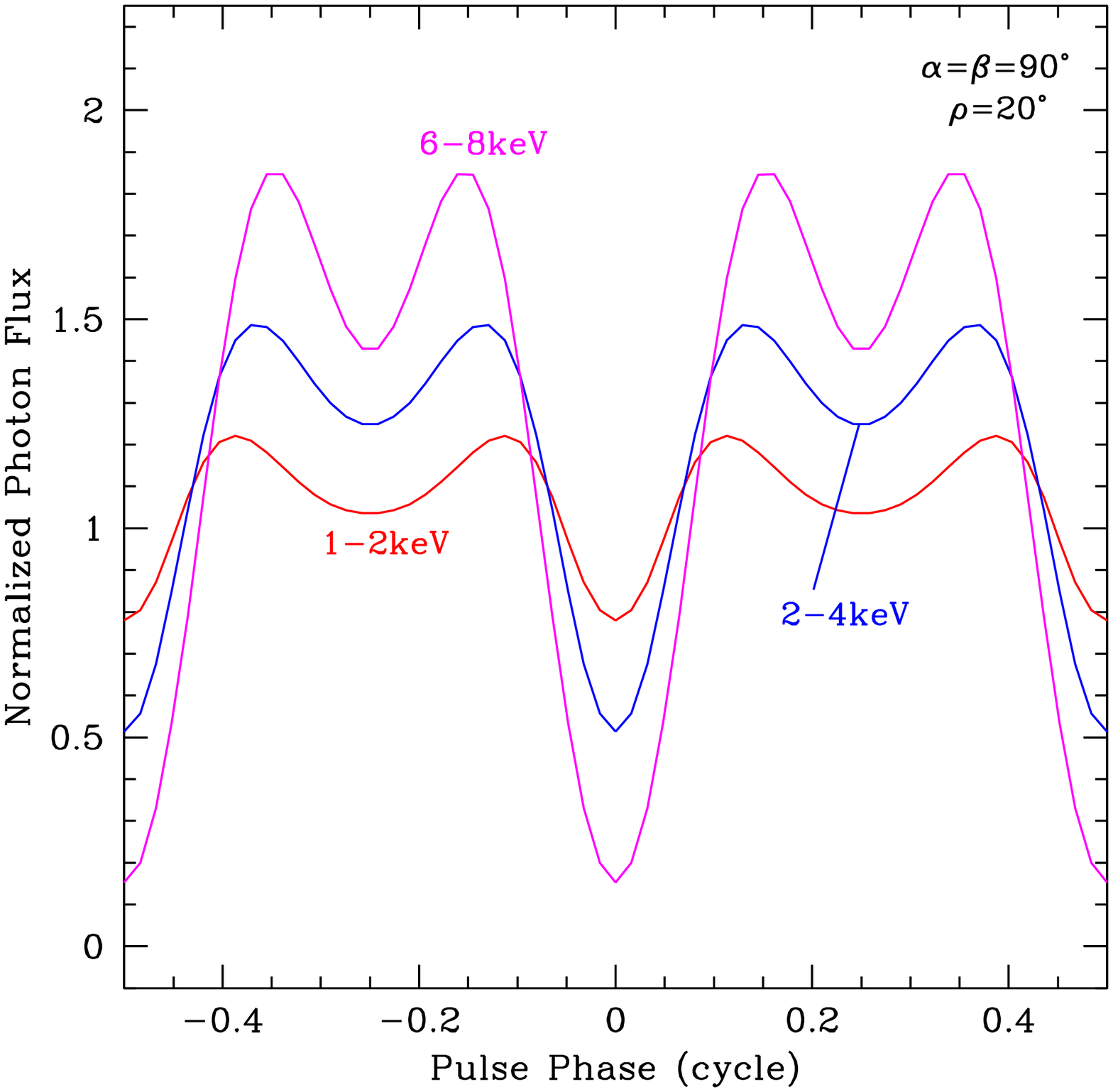,angle=0,width=8.2truecm} }
\figcaption[]{Pulse profiles that have rather similar morphologies in
different energy bands between 1-8 keV. The top panel shows the
results for a single pole while the bottom panel shows the results for
an antipodal emission geometry. \label{Fig:pp_e}}
\end{figure}

Figure~\ref{Fig:pp_e} shows the case of an orthogonal rotator with one
and two $20^\circ$ hot poles (top and bottom panels respectively) on a
neutron star with compactness $p=2.5$. For this set of parameters, the
energy dependence in the single pole case is quite weak and is
characterized by a small phase shift in the location of the peak of
the pulse profiles between different energies. In addition, there is a
more prominent interpulse (at $\phi=0$ in Fig.~\ref{Fig:pp_e}) at low
photon energies. The pulse amplitudes are nearly equal in the three
bands. This type of weak dependence on photon energy is very common in
the single-pole geometry and is obtained for most of the model
parameters.

The slight phase shift and broader pulses are a direct consequence of
the different beaming of radiation at different photon energies. As
shown earlier in Figure~\ref{Fig:beaming}, the peak of the fan beam
appears at smaller angles $\theta_{\rm p}$ for smaller photon
energies, giving rise to narrower emission cones. Furthermore, at
these lower energies, the beams are flatter and the minimum separating
the fan and pencil beams becomes less prominent. Thus, for the
orthogonal rotator geometry, where an observer intersects the emission
cones at the diameter, the lower energy peaks appear more closely
spaced than the higher energy peaks and the flux between them is
higher.

\begin{figure}[t]
\centerline{ \psfig{file=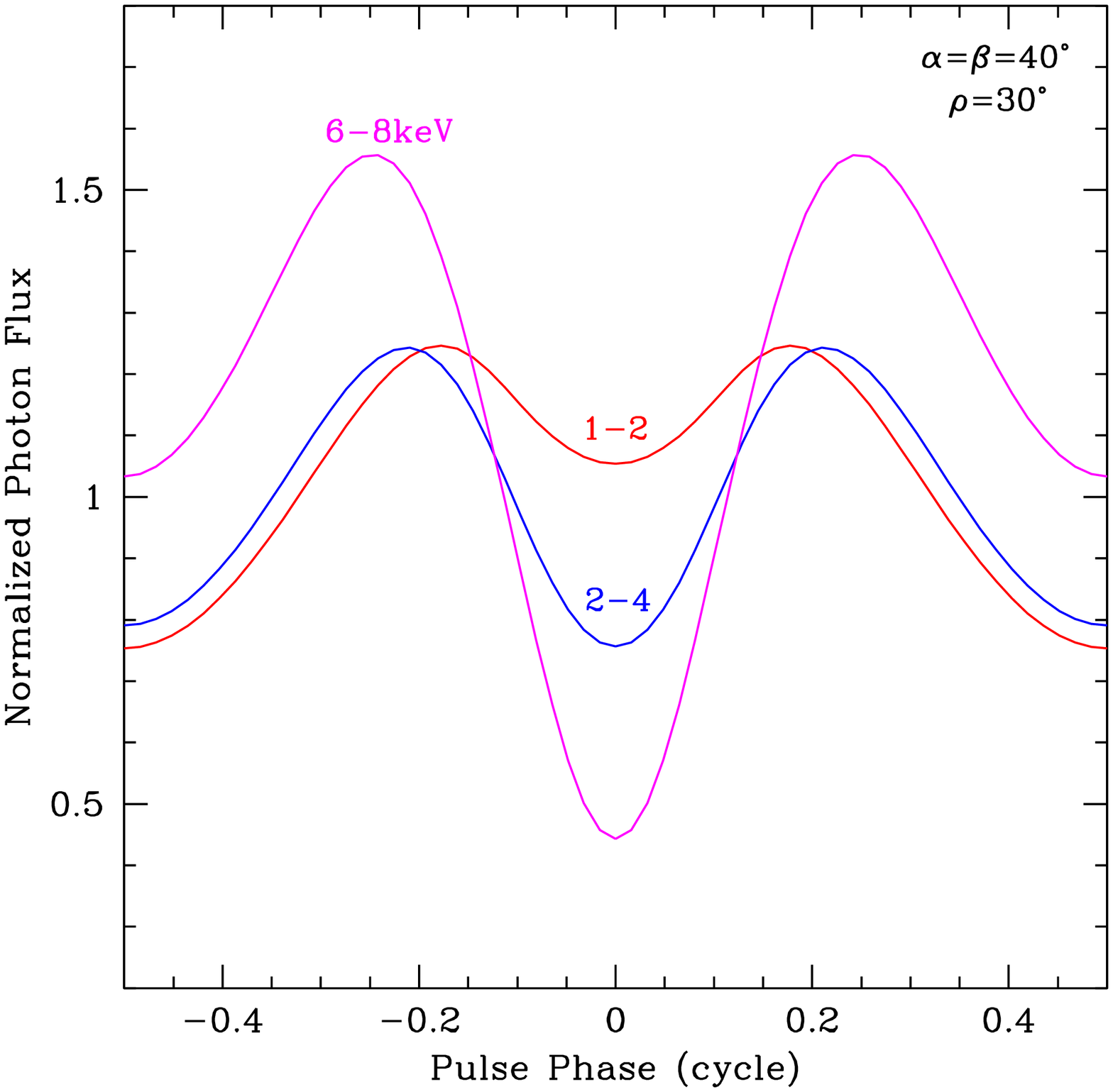,angle=0,width=8.2truecm} }
\centerline{ \psfig{file=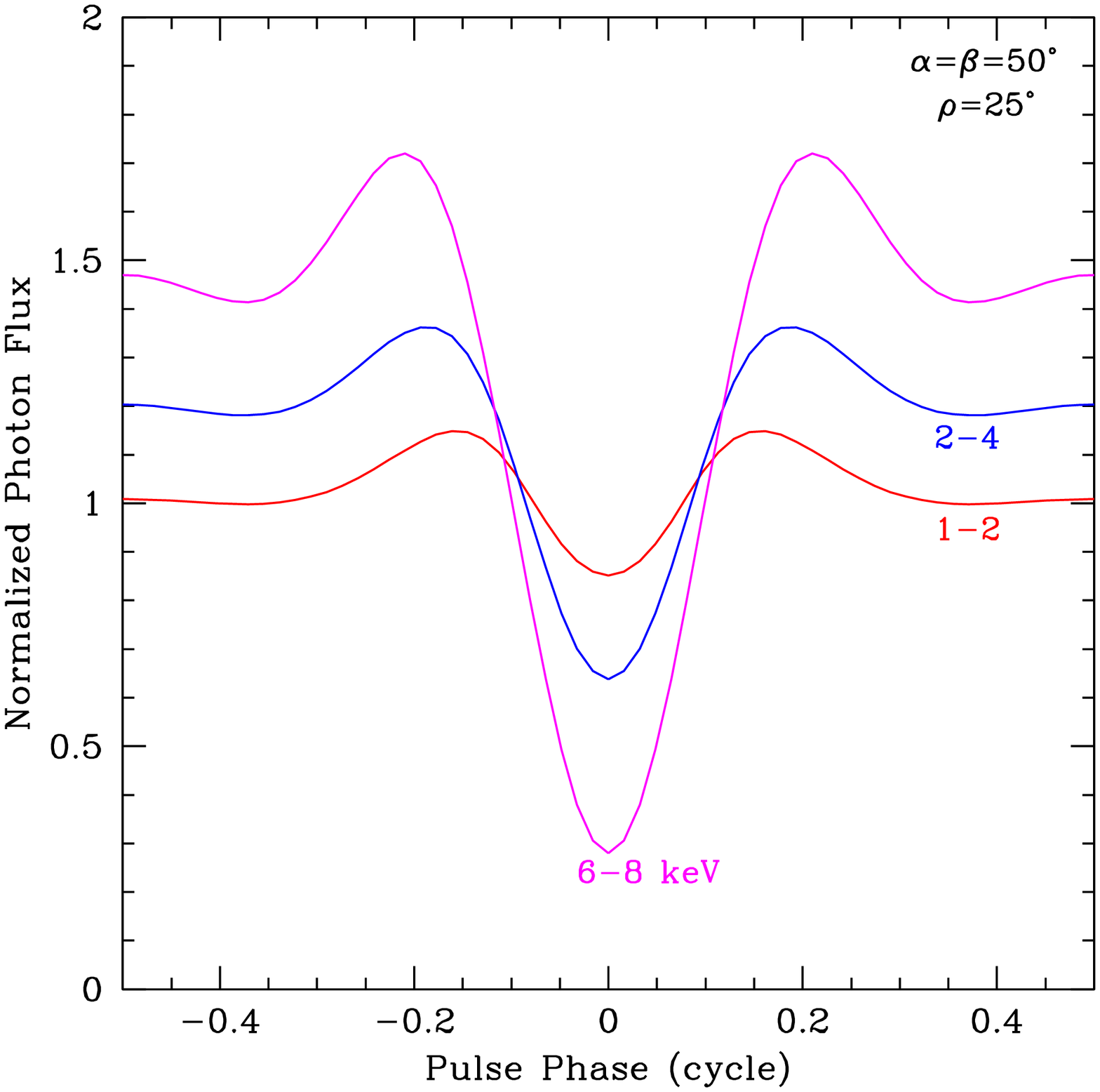,angle=0,width=8.2truecm} }
\figcaption[]{A strong photon-energy dependence of the pulse profiles
between 1-8 keV. The top panel corresponds to a single and the bottom
panel to two antipodal hot regions on the neutron star
surface. \label{Fig:pp_estrong}}
\end{figure}

In the antipodal emission geometry, the pulse shapes in the different
energy bands are also similar and the small phase shift in the
location of the peaks is present. On the other hand, there is a marked
difference in the pulse amplitudes in the different energy bands. Note
that Figure~7 shows one of the weakest energy dependences in the pulse
morphology that can be obtained for any choice of parameters in this
emission geometry.

Figure~\ref{Fig:pp_estrong} shows energy-dependent pulse profiles for
different values of the observer's angle $\beta$ and the angular size
of the emitting area $\rho$. In these examples, the energy dependence
is stronger, both in the pulse shapes and in the pulse amplitudes.  In
the case of one pole, the cross-correlation of the $1-2$ and $6-8$~keV
pulse profiles reveals an apparent phase lag of half a cycle caused
primarily by the difference in the phase of the minima in the two
profiles.  In the antipodal geometry, the number of peaks increases by
one in the higher photon-energy band. In both cases, the pulse
amplitude is highest in the $6-8$~keV band. Such simultaneous changes
in both the pulse morphology and in the pulse amplitude are rare in
the case of a single pole but are obtained for most of the parameter
space for the antipodal geometry.

The bottom panel of Figure~\ref{Fig:pp_estrong} along with that of
Figure~\ref{Fig:pp_e} spans the entire range of energy-dependent pulse
profiles that can arise in the antipodal emission geometry. In the
case of a single hot pole, on the other hand, the energy dependence of
the pulse profiles shows more diversity, as the following examples
illustrate.  

\begin{figure}[t]
\centerline{ \psfig{file=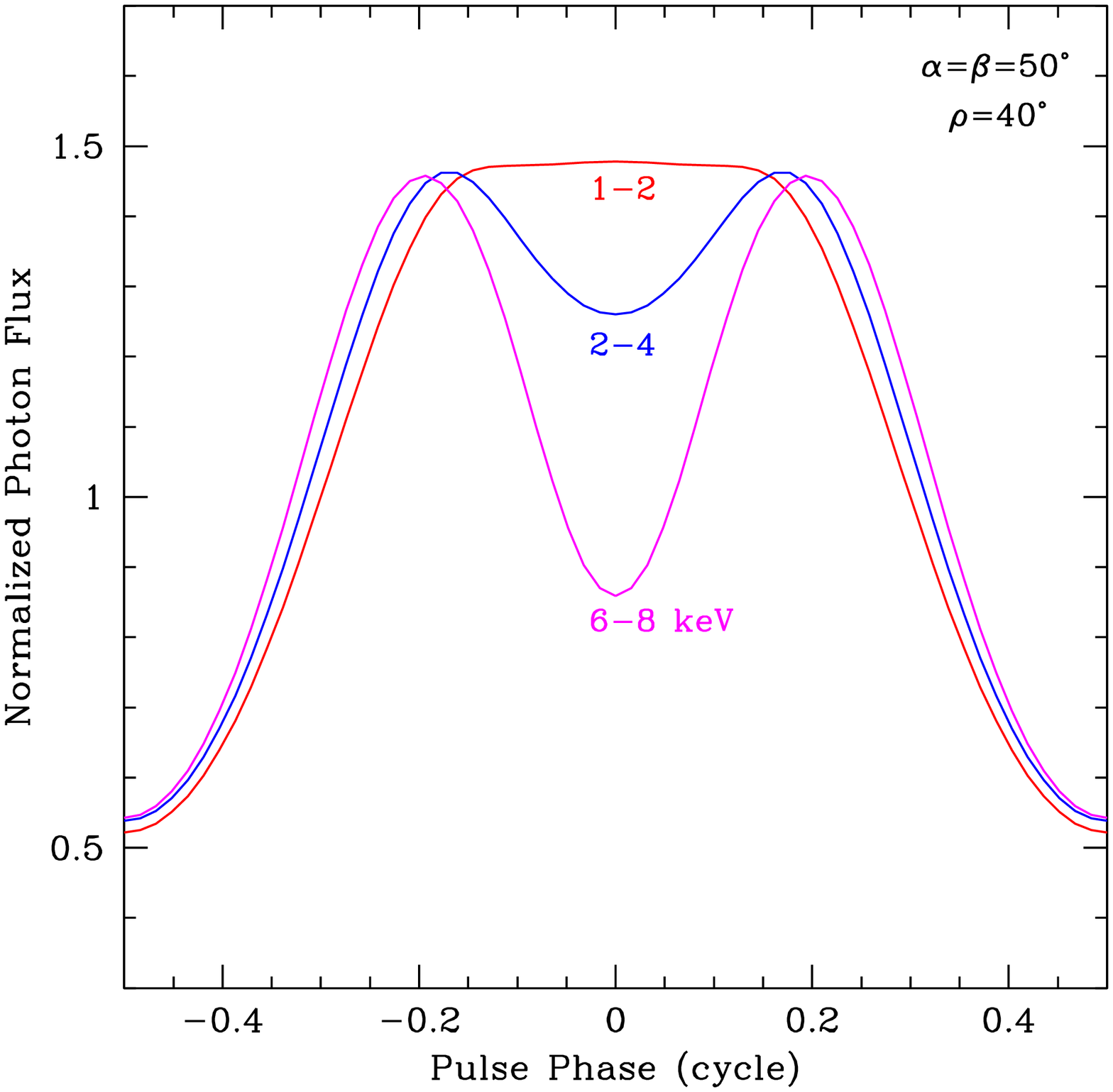,angle=0,width=8.5truecm} }
\figcaption[]{An example of a pulse profile at different photon
energies from a magnetar with a single hot pole, where the pulse
shapes are energy-dependent while the pulse amplitudes remain
constant. \label{Fig:pp_eweak}}
\centerline{ \psfig{file=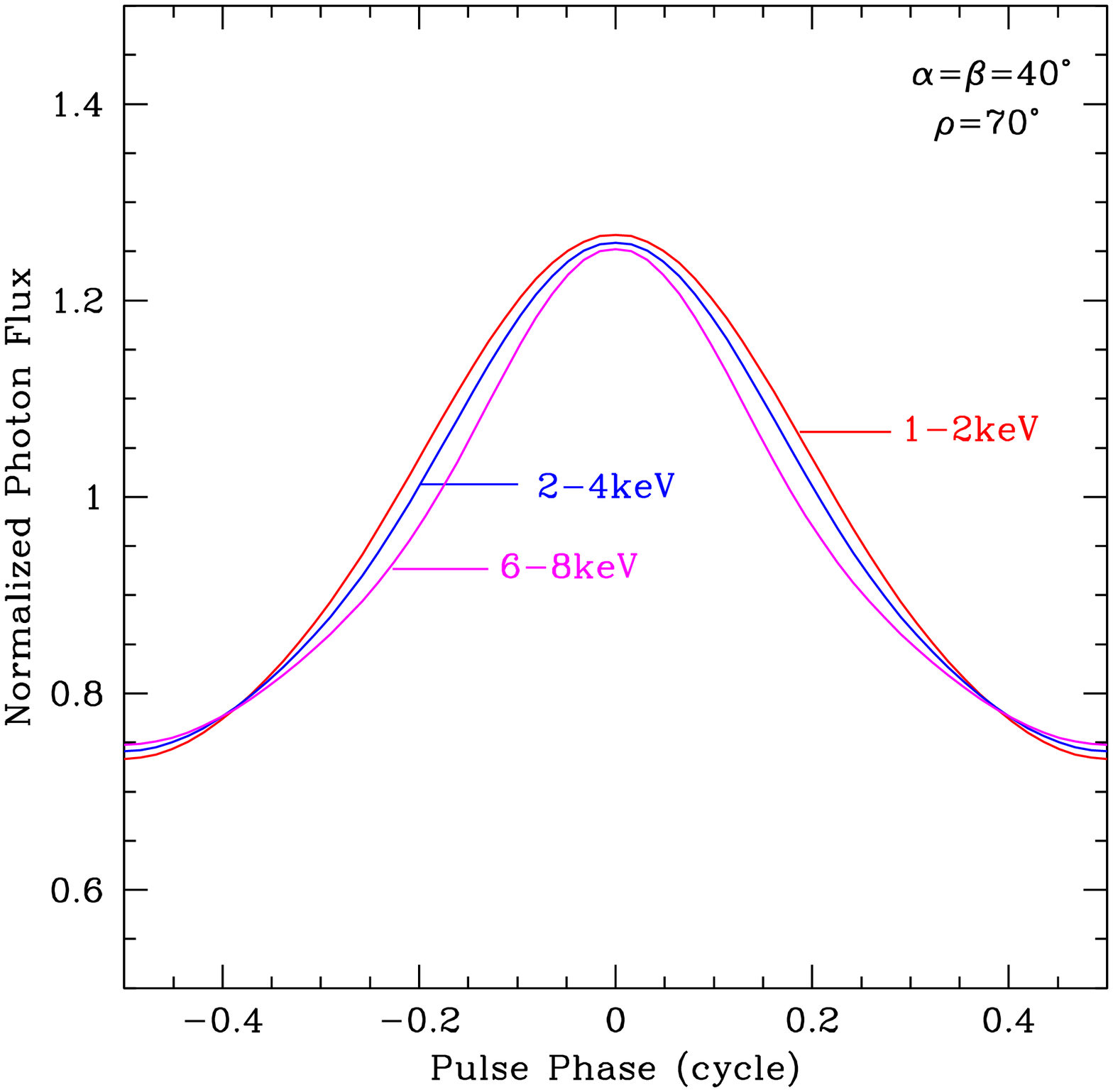,angle=0,width=8.5truecm} }
\figcaption[]{An example of a singly-peaked pulse profile at different
photon energies from a magnetar with a single hot pole, where the
pulse shapes and amplitudes show no energy
dependence. \label{Fig:pp_eone}}
\end{figure}

Figure~\ref{Fig:pp_eweak} shows a configuration where the
pulse morphology, and in particular the number of peaks, changes with
photon energy, while the pulse amplitude remains constant.  Finally,
Figure~\ref{Fig:pp_eone} shows a pulse profile with a single peak and
no energy dependence. This is obtained when the observer grazes the
emission cones at all photon energies.  Such a case is relevant to the
observations of 1E~$1048.1-5937$ which is the only AXP with a
singly-peaked pulse profile and shows no energy dependence (Kaspi et
al.\ 2001; Tiengo et al.\ 2002).

\vspace{-0.3cm}
\section{Pulsed Fractions}

In this section, we quantify the modulations in the pulse profiles
using the pulsed fraction defined by 
\be 
PF = \frac{F_{\rm max}-F_{\rm min}}{F_{\rm max}+F_{\rm min}}, 
\ee 
where $F_{\rm max}$ and $F_{\rm min}$ are the maximum and minimum
fluxes received by an observer at infinity during a pulse cycle.
Bolometric pulsed fractions have been discussed in DeDeo, Psaltis, \&
Narayan (2001) for a parametrized form of radially peaked beaming
functions, which are not directly applicable to thermally emitting
magnetars.  Here we present results using the beaming functions
calculated for the ultramagnetized neutron stars as well as
investigate the dependence of the pulsed fractions on various model
parameters. (See also \"Ozel et al.\ 2001 for an earlier discussion).

\begin{figure}[t]
\centerline{ \psfig{file=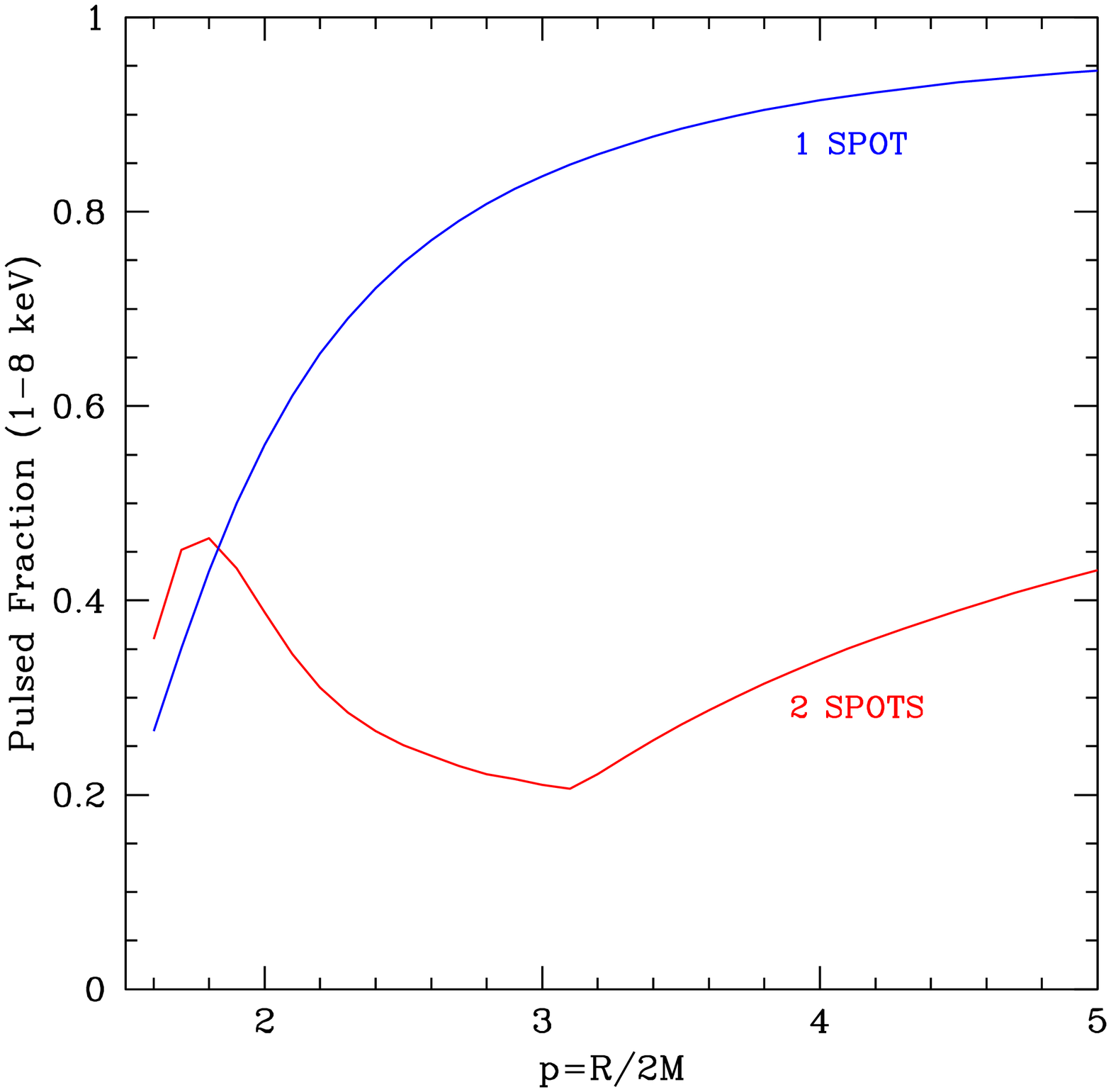,angle=0,width=8.5truecm} }
\figcaption[]{The dependence of the $1-8$~keV pulsed fraction on the
compactness of the neutron star for a single and two antipodal hot
regions. In the single pole case, $\alpha=\beta=60^\circ$ and
$\rho=40^\circ$, while in the antipodal emission geometry,
$\alpha=\beta=90^\circ$ and $\rho=25^\circ$. \label{Fig:pf_p}}
\centerline{ \psfig{file=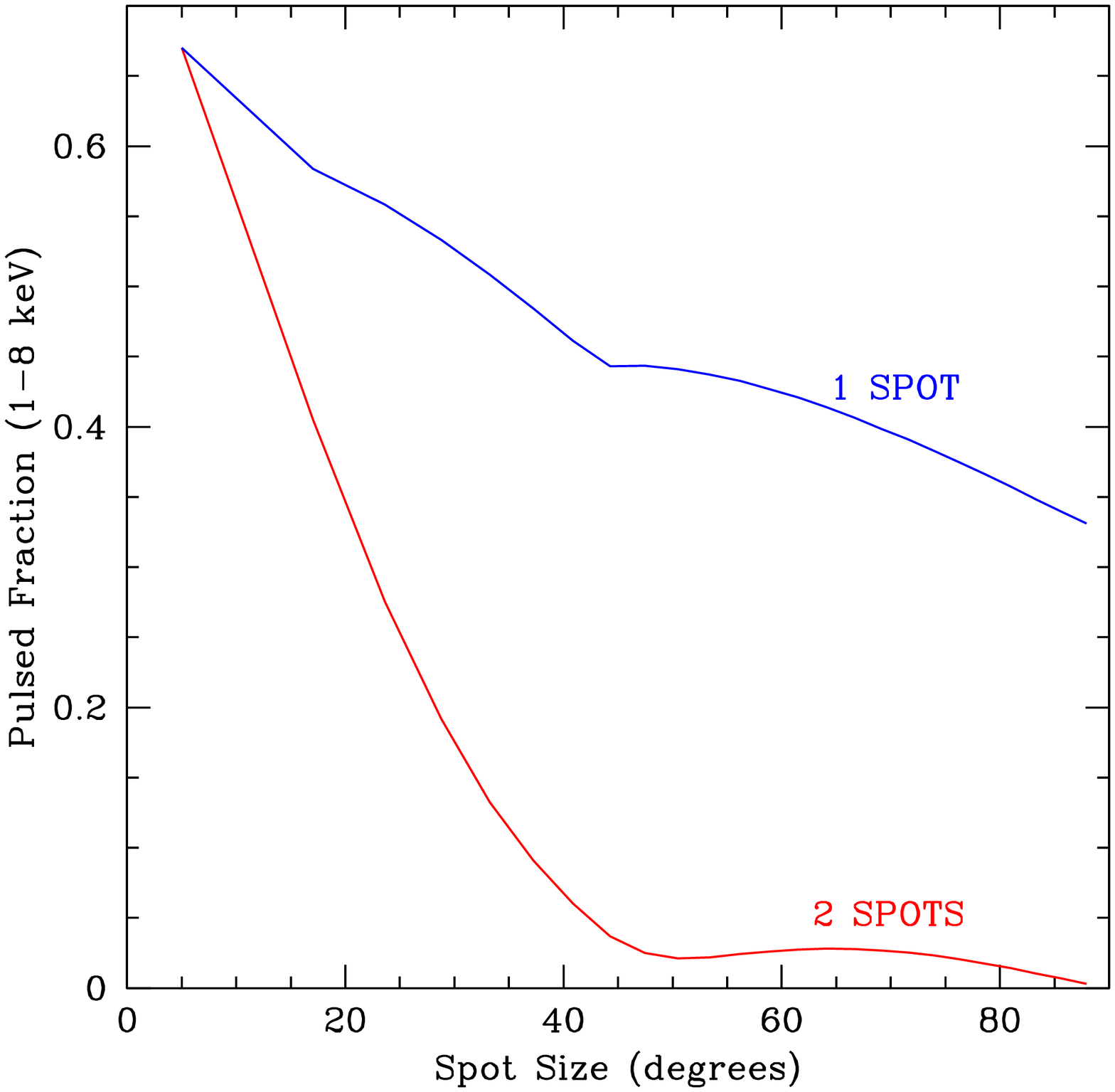,angle=0,width=8.5truecm} }
\figcaption[]{The dependence of the pulsed fraction on the size of the
emitting region for $p=2.5$ and $\alpha=\beta=50^\circ$. Because of
gravitational lensing effects, the dependence is
non-monotonic. \label{Fig:pf_rho}}
\end{figure}

Figure~\ref{Fig:pf_p} shows the dependence of the bolometric pulsed
fraction on the relativity parameter $p$ of the neutron star for the
case of one and two poles. General relativistic effects alter the
observable modulation of X-ray emission originating from a neutron
star surface and in general lead to a significant suppression of the
pulse amplitude when the gravitational lensing by the star is strong.
Therefore, the pulsed fraction is expected to decrease monotonically
as the neutron star becomes more relativistic. This is indeed the case
seen in the single pole geometry.

The antipodal emission case, on the other hand, shows a more complex
behavior. As $p$ decreases, we first obtain the more commonly observed
suppression of the pulse amplitudes. A qualitative change occurs at $p
\approx 3$ when the lensing by the neutron star is strong enough to
lead to the previously discussed enhancement of the flux at the
intermediate phases $\phi=\pm \pi/2$, at which the emission cones
intersect (see Fig.~\ref{Fig:ppbol_p}). This causes the pulsed
fraction to increase towards smaller relativity parameters. As $p$
decreases further, nearly the entire surface becomes visible to the
observer and thus the behavior is again reversed.

Figure~\ref{Fig:pf_rho} shows the dependence of the pulsed fraction on
the size of the emitting region for the two surface temperature
distributions. This figure serves as a typical example and summarizes
the two important aspects of how the pulsed fraction changes with
size: (i) The pulsed fraction falls off much more slowly with
increasing angular size of a single pole compared to that of two
antipodal hot regions. This results in a much higher pulsed fraction
for the same total luminosity (for a fixed effective temperature) in
the single pole case since luminosity grows linearly with surface
area. As discussed in \"Ozel et al.\ (2001), this has important
implications for the AXPs, which have both high pulsed fractions and
high inferred luminosities. (ii) The pulsed fraction varies
non-monotonically with increasing size of the hot regions. This
follows directly from the discussion in \S 3.2, where we showed that
the pulse profiles go through a sudden change at $\rho \sim 40^\circ$
when the non-radial beams from the outer edges of a large hot region
are seen simultaneously by an observer, resulting in an enhanced flux.
The pulsed fraction, therefore, goes through a corresponding increase
when the peak of the pulse profile shifts to $\phi=0^\circ$.

\begin{figure}[h]
\centerline{ \psfig{file=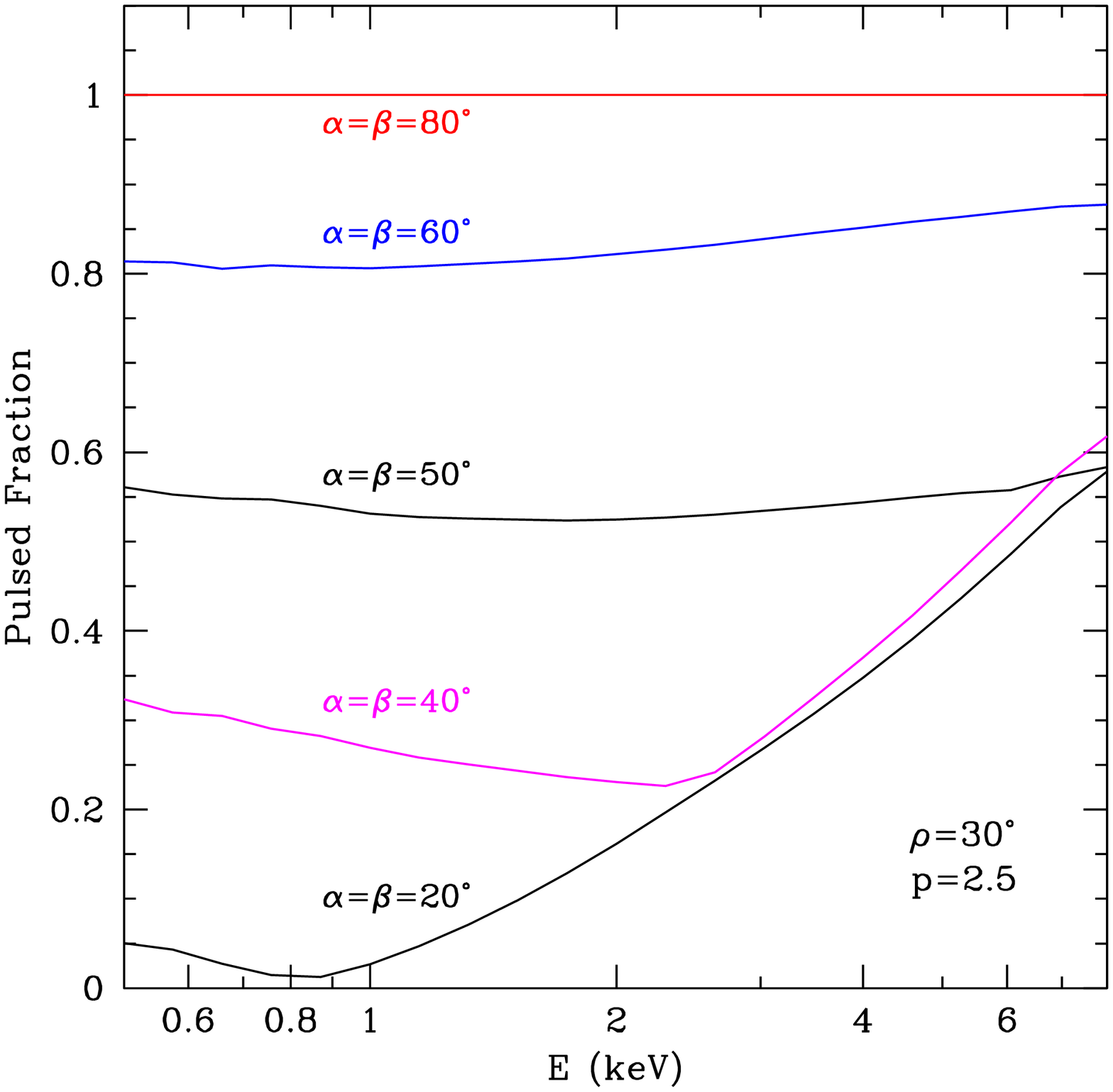,angle=0,width=8.0truecm} }
\centerline{ \psfig{file=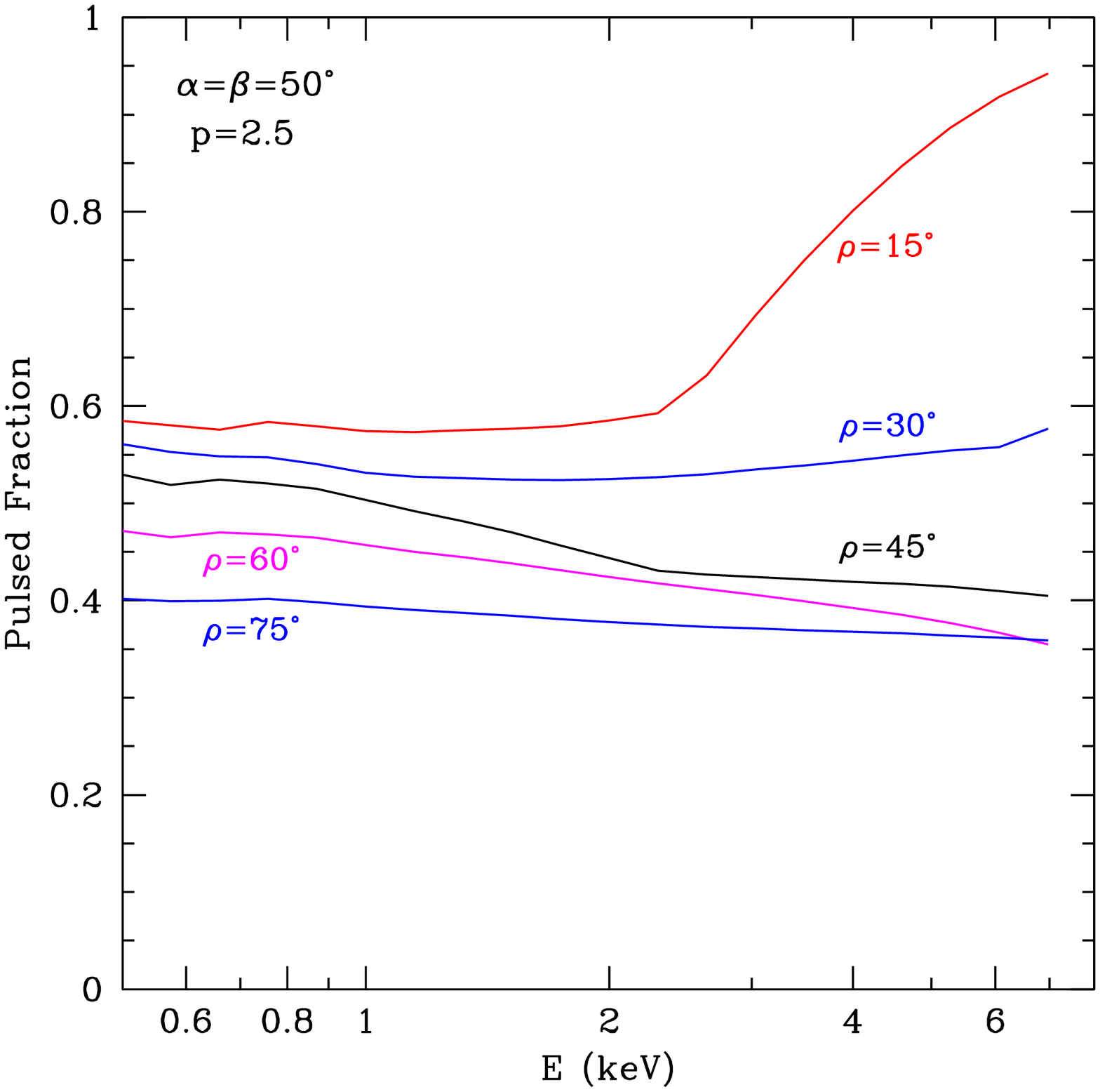,angle=0,width=8.0truecm} }
\figcaption[]{The dependence of the pulsed fraction on photon energy
for (top panel) a range of observer angles and (bottom panel) a range
of sizes of the emitting area when the emission originates from a
single hot pole. For most of the parameter space, the dependence is
very weak.\label{Fig:pf_e1}}
\end{figure}

Finally, Figure~\ref{Fig:pf_e1} shows, for the single pole case, the
dependence of the pulsed fraction on photon energy for a range of
orientation angles $\alpha$ and $\beta$ and for different angular
sizes of the hot region $\rho$.  For $\rho \gtrsim 20^\circ$ and
$\alpha$ or $\beta \gtrsim 50^\circ$, the pulsed fractions are
remarkably constant within the energy range of interest. It is
interesting to note that these ranges of parameters cover most of the
parameter space of orientation angles for a random distribution and
correspond at the same time to the brightest sources because of their
large emitting areas.  For the other values of the parameters, the
pulsed fractions show an increase with photon energy that is also
characteristic of the antipodal emission geometry for nearly all
parameter values (see \"Ozel et al.\ 2001 for the corresponding figure
in the antipodal emission geometry and a discussion of the physical
reasons behind this characteristic shape).

\section{Discussion}

We studied the timing properties of ultramagnetized neutron stars
emitting thermally from their surfaces. We used the energy and angle
dependence of the emerging radiation obtained from the recent detailed
calculations of radiative equilibrium atmospheres in ultrastrong
magnetic fields (\"Ozel 2001, 2002). We combined them with the general
relativistic calculation of photon transport from the neutron star
surface to an observer at infinity and calculated the expected pulse
profiles for a wide range of model parameters. We found that the
combination of the non-radially peaked beams relevant to magnetars
with the strong gravitational lensing leads to a number of
qualitatively new and interesting results on the pulse morphology and
amplitudes, which we summarize below.

\noindent {\bf 1.} An emission geometry consisting of one hot pole
gives rise to one or two peaks per pulse cycle, whereas an antipodal
geometry can produce one to four peaks, depending strongly on the
orientation angles $\alpha$ and $\beta$ as well as on the relativity
parameter $p$ and the angular size of the emitting region $\rho$.

\noindent {\bf 2.} The pulse profiles in different energy bands
show a wide range of morphologies in both emission geometries.

\noindent {\bf 3.} The non-radial beaming of the thermal radiation
emerging from an ultramagnetized neutron-star surface can give rise
to high pulsed fractions even for very compact neutron stars.

\noindent {\bf 4.} In the case of an antipodal emission geometry, the
pulsed fraction may not be a monotonic function of the relativity
parameter $p$, in contrast to the case of radially peaked beaming
patterns.

\noindent {\bf 5.} The pulsed fraction does not decrease monotonically
with the size of the emitting region but shows a secondary maximum at
intermediate ($\sim 60-70^\circ$) angular sizes.

\noindent {\bf 6.} The pulsed fraction in the antipodal emission
geometry shows a characteristic increase with photon energy in the
1--8~keV range. In contrast, the pulsed fraction in the single
pole case has, in general, a very weak energy dependence. 

Note that a number of simplifying assumptions have been made in the
models presented here. We have taken the temperature across a hot
region to be uniform, which may not be realistic for all the
mechanisms that could power the surface emission. In the case of a
cooling neutron star with a dipole magnetic field, the temperature is
a function of the magnetic latitude and thus shows variations across
the surface of the neutron star (e.g., Page 1995). Also in the case of
a single hot pole, similar variations can also be expected. This can
affect the pulse profiles, possibly reducing the pulsed fractions and
smoothing out the pulse morphologies. Similarly, allowing for some
cooler emission from the rest of the neutron star can alter the pulse
profiles and may account for the additional complexity observed in the
energy-dependent pulse profiles of AXPs.

We also assumed that the emission properties are determined entirely
by the stellar atmosphere and have not taken into account the possible
effects of the neutron star's magnetosphere further above (see, e.g.,
Thompson, Lyutikov, \& Kulkarni 2002).  At present, the structure of
the magnetosphere of a strongly magnetized neutron star is not well
understood and can only be treated through parametrizations. Note that
the alternative class of models of AXPs that rely on accretion onto a
magnetized neutron star also require the study of the processes that
take place in the accretion column in the neutron star's magnetic
field. It was shown earlier that such processes can also lead to a
variety of pulse profiles (e.g., M\'esz\'aros \& Nagel 1985) but further
study is necessary to determine their relevance for AXPs.

The results presented above have direct implications for the thermally
emitting magnetar models of AXPs. To carry out a comparison with
observations, we summarize the timing properties of these sources
which can be used to constrain such models.  First, the observed pulse
profiles show two prominent peaks per pulse cycle in four sources and
a single peak in the fifth source (e.g., Gavriil \& Kaspi
2002). Second, the energy dependence of the pulse morphology can be
very weak, as in 1E~1048.1$-$5937, or quite strong, as in
1RXS~1708$-$4009 (Gavriil \& Kaspi 2002). Third, AXPs can have pulsed
fractions as high as 70\%, together with luminosities that are high
for their low inferred effective temperatures (\"Ozel et al.\
2001). Finally, their pulsed fractions show a weak dependence on
photon energy (e.g., Oosterbroek et al.\ 1998).

All the above properties are hard to account for in a thermally
cooling magnetar model (e.g., Heyl \& Hernquist 1998) that has a
two-fold symmetry and thus has pulse profiles very similar to those of
the antipodal emission geometry. In particular, the number of observed
peaks in the pulse profiles as well as the strong energy-dependence
predicted in the two-pole models do not reproduce the observations.
Furthermore, the pulsed fractions in such a model are in general lower
than the values observed in AXPs. 

The observations, on the other hand, strongly suggest that AXPs are
neutron stars with a single, hot region of ultrastrong magnetic
field. A single hot pole can be realized in a number of ways. It may
be powered by processes such as crustal cracking, magnetic-field
reconfiguration, or decay of magnetic multipoles (e.g., Thompson \&
Duncan 1996). It could also arise from an off-centered magnetic dipole
that renders one of the poles either not observable or else very close
to the other pole.  All of these possibilities need further
theoretical investigations in order to determine the geometry of
emission, the total energy output and the lifetime of the temperature
asymmetries on the surface.  In addition, because the first set of
processes are expected to occur randomly, they can produce
time-variable pulse profiles. Observations of pulse morphology changes
similar to the {\em GINGA} observation of 1E~2259$+$586 (Iwasawa,
Koyama, \& Halpern 1992) can help constrain the mechanism powering the
thermal emission of AXPs. Finally, investigating the energy dependence
of the pulsed fractions in the soft X-rays ($\lesssim 1$~keV) with
{\em Chandra\/} and {\em XMM-Newton\/} as well as in longer
wavelengths will extend the baseline over which models can be compared
to data and provide the most stringent constraints.

\acknowledgements

I am grateful to Fotis Gavriil and Vicky Kaspi for sharing with me
their AXP data prior to publication and for many stimulating
discussions on pulse profiles. I thank Dimitrios Psaltis and Ramesh
Narayan for useful suggestions and discussions. I also thank the
Institute for Advanced Study, where this work was completed, for their
hospitality.  This work was supported in part by the NASA Chandra
grant GO0-1105B and by the NSF grant AST 9820686.


\clearpage


\begin{references}

\reference{2001ApJ...554.1245A} Alpar, M.~A.\ 2001, \apj, 554, 1245

\reference{Cardall} Cardall, C.~Y., Prakash, M., \& Lattimer, J.~M.\
2001, \apj, 554, 322

\reference{2000ApJ...534..373C} Chatterjee, P., Hernquist, L., \&
Narayan, R.\ 2000, \apj, 534, 373

\reference{Dedeo} DeDeo, S., Psaltis, D., \& Narayan, R.\ 2001, 
\apj, 559, 346

\reference{Gavriil} Gavriil, F. \& Kaspi, V.\ 2002, \apj, in press

\reference{HH98} Heyl, J.\ \& Hernquist, L.\,E.\ 1998, \mnras, 300, 599

\reference{2001MNRAS.327.1081H} Ho, W.~C.~G.~\& Lai, D.\ 2001, 
\mnras, 327, 1081

\reference{Iwasawa} Iwasawa, K., Koyama, K., \& Halpern, J.\,P.\ 
1992, \pasj, 44, 9

\reference{Juett} Juett, A.\ M., Marshall, H.\ L., Chakrabarty, D.,
Canizares, C.\ R., Schulz, N.\ S.\ 2002, in Neutron Stars in Supernova
Remnants, eds P.\ O.\ Slane and B.\ M.\ Gaensler (ASP Conference
Proceedings)

\reference{kaspietal} Kaspi, V.~M., Gavriil, F.~P., Chakrabarty, D.,
Lackey, J.~R., \& Muno, M.~P.\ 2001, \apj, 558, 253

\reference{mn} M\'esz\'aros, P.~\& Nagel, W.\ 1985, \apj, 299, 138. 

\reference{Oetal98} Oosterbroek, T., Parmar, A.\,N., Mereghetti, S.,
\& Israel, G.\,L.\ 1998, A\&A, 334, 925

\reference{2001ApJ...563..276O} \"Ozel, F.\ 2001, \apj, 563, 276

\reference{ozel02} \"Ozel, F.\ 2002, \apj, submitted

\reference{2001ApJ...563..255O} \"Ozel, F., Psaltis, D., \& Kaspi, V.~M.\ 
2001, \apj, 563, 255

\reference{page} Page, D.\ 1995, \apj, 442, 273

\reference{2001ApJ...563L..45P} Patel, S.~K.~et al.\ 2001, \apjl, 563, L45

\reference{Tiengo et al} Tiengo, A., G\"ohler, E., Staubert, R., \& 
Mereghetti, S.\ 2002, A\&A, in press (astro-ph/0111304)

\reference{1995A&A...299L..41V} 
van Paradijs, J., Taam, R.~E., \& van den Heuvel, E.~P.~J.\ 1995, 
\aap, 299, L41

\reference{TD96} Thompson, C.\ \& Duncan, R.\ C.\ 1996, \apj, 473, 322

\reference{TLK} Thompson, C., Lyutikov, M., \& Kulkarni, S.\ 2002,
\apj, submitted (astro-ph/0110677)

\reference{TE} Tsai, W. \& Erber, T.\ 1975, \prd, 12, 1132

\reference{1996ApJ...463L..83W} White, N.\,E., Angelini, L., Ebisawa, K., 
Tanaka, Y., \& Ghosh, P.\ 1996, \apjl, 463, L83

\reference{2001ApJ...560..384Z} Zane, S., Turolla, R., Stella, L., \& 
Treves, A.\ 2001, \apj, 560, 384

\end{references}
\end{document}